\documentclass[final,1p,times]{elsarticle}%
\usepackage{amssymb}
\usepackage{amsmath}%
\usepackage{amsfonts}%
\usepackage{graphicx}
\journal{}
\begin{document}
\begin{frontmatter}
\title{Efficient and accurate computation of electric field dyadic Green's function in layered media}
\author[1]{Min Hyung Cho}
\author[2]{Wei Cai}
\address[1]{Department of Mathematical Sciences, University of Massachusetts Lowell, Lowell MA 01854}
\address[2]{Department of Mathematics and Statistics, University of North Carolina at Charlotte, Charlotte NC 28223}
\begin{abstract}
Concise and explicit formulas for dyadic Green's functions, representing the electric and magnetic fields due to a dipole source placed in layered media, are derived in this paper. First, the electric and magnetic fields in the spectral domain for the half space are expressed using Fresnel reflection and transmission coefficients. Each component of electric field in the spectral domain constitutes the spectral Green's function in layered media. The Green's function in the spatial domain is then recovered involving Sommerfeld integrals for each component in the spectral domain. By using Bessel identities, the number of Sommerfeld integrals are reduced, resulting in much simpler and more efficient formulas for numerical implementation compared with previous results. This approach is extended to the three-layer Green's function. In addition, the singular part of the Green's function is naturally separated out so that integral equation methods developed for free space Green's functions can be used with minimal modification. Numerical results are included to show efficiency and accuracy of the derived formulas.
\end{abstract}
\end{frontmatter}



\section{Introduction}

Multi-layered media is a fundamental structure for many applications such as
meta-materials, photonic crystals \cite{jobook}, solar cells \cite{atwater,
qd_solar}, light emitting diodes \cite{led}, and plasmonic devices and others.
Numerical simulation of wave propagation in such media poses much challenge
due to large number of scatters, the treatment of radiation condition at the
infinite, and the field discontinuity at layer interfaces in meta-materials
consisting of meta-atoms. Integral equation methods have been shown to be
versatile to address these issues in computing the wave scattering in the
layered media. To implement the integral equation formulation of the
scattering problem, it is imperative to have a concise formulation and
efficient computational algorithm to compute the dyadic Green's functions for
the Maxwell's equations in the three-dimension (3-D). In this paper, we will
present explicit and compact formulas for the two- and three-layer dyadic
Green's functions in terms of high order Hankel transforms and relevant
numerical method for their computations.

The dyadic Green's function for a two-layer structure \cite{chew1999} and
multi-layered media \cite{cui1998} have been explicitly presented. However,
the formula for the three or more layers was not provided in Ref.
\cite{chew1999}. Also, the derivations in these work used an analytical
formula for Sommerfeld integrals for two layers in order to reduce the total
number of Sommerfeld integrals to 10. As a consequence, extension to multi-layered media for
sources on top of the layered media as well as in the middle layer is not
obvious. As a result, multi-layered media Green's function requires extra
Sommerfeld integrals. The multi-layered media Green's function in
\cite{cui1998} requires total 16 Sommerfeld integrals. The new formula
proposed in this paper utilizes the second order Hankel transform to reduce
the number of integrals needed and the singular and nonsingular parts of the
Green's function are clearly separated. This allows easy use of many integral equation algorithms and codes
developed using free space Green's function  \cite{vie_jcp} or periodizing
schemes for periodic objects \cite{helmholtz_periodic, junlai} for the
multi-layered media problems. Moreover, our approach in principle,
with some more bookkeeping associated with the layers, can be extended to the
multi-layered media when the source is on top of the layered media. Discussion on various numerical 
issues of implementing the integral equations can be found in Ref. \cite{caibook,
algorithmic_issue} and it is not repeated here. For numerical contour
integration of Sommerfeld integrals in the Fourier $k$-space, adaptive
generalized Gaussian quadrature rules \cite{gen_quad, chocai12} are used to
obtain high accuracy using quadrature points only on the real axis. This
avoids complex number operations and reduces computation time. In other words,
near the surface poles of the spectral dyadic Green's functions, generalized
Gaussian quadrature rule is applied while traditional Gaussian quadrature is
applied in other parts of the contour.

The derivation for the dyadic Green's function in this paper is rather
cumbersome and tedious. However, it is unavoidable for multi-layered media
simulation and much needed in practice of integral equations using dyadic
Green's functions. Every effort is made to simplify the final formula so the
readers can implement them easily. The same notation as in Ref.
\cite{chew1999} will be used and modified as necessary throughout the paper.

The rest of the paper is organized as follows. In the next section, the
free-space Green's function is transformed to one in the spectral domain using
the Sommerfeld identity. Then, the two-layer Green's functions will be derived
using the free-space Green's function and Fresnel reflection coefficients
\cite{stratton, yeh} in Section 3. In Section 4, extension will be given for
the three-layer Green's functions with generalized Fresnel reflection
coefficients \cite{chewbook} due to multiple reflections from the interfaces.
Finally, in Appendix, several Bessel identities used for the derivations are provided.

\section{Free-space Green's function}

The free-space Green's function serves as a primary singular field for the
multi-layered media Green's function. In multi-layered media, the free-space
Green's function will be ``corrected" with reflected and transmitted
contribution. Thus, in this section, the dyadic Green's function for the free space is studied. First, it is rewritten in the spectral domain. Then, the
spatial domain Green's function is recovered by taking the inverse Fourier
transform. The same process will be applied for multi-layered media.
For convenience, the free space will be referred as a one-layer problem that has relative permittivity $\varepsilon_{1}$ and permeability $\mu_{1}$. Let a unit dipole be placed at $\mathbf{r}^{\prime}= (x^{\prime},y^{\prime}, z^{\prime})$ and oriented along $\hat{\boldsymbol{\alpha}}^{\prime}= (\alpha^{\prime}_{x},\alpha^{\prime}_{y},\alpha^{\prime}_{z})$. Then, the electric $\mathbf{E}_{1}^{P} = (E_{1x}^{P}, E_{1y}^{P}, E_{1z}^{P})$ and magnetic $\mathbf{H}_{1}^{P} = (H_{1x}^{P}, H_{1y}^{P}, H_{1z}^{P})$ fields in the free space at $\mathbf{r} = (x,y,z)$ can be written as
\begin{align}
\mathbf{E}_{1}^{P}(\mathbf{r})  & = \frac{i\omega\mu_{0}\mu_{1} }{4\pi}
(\mathbf{I}+\frac{\nabla\nabla}{k^{2}_{1}}) \cdot\hat{\alpha}^{\prime}%
\frac{e^{ik_{1}|\mathbf{r}-\mathbf{r}^{\prime}|}}{|\mathbf{r}-\mathbf{r}%
^{\prime}|},~~~~ \mathbf{H}_{1}^{P}(\mathbf{r}) = \frac{1}{4\pi} \nabla
\times\hat{\alpha}^{\prime}\frac{e^{ik_{1}|\mathbf{r}-\mathbf{r}^{\prime}|}%
}{|\mathbf{r}-\mathbf{r}^{\prime}|},\label{primary_h}%
\end{align}
where $k_{1} = k\sqrt{\varepsilon_{1} \mu_{1}}$ in the dielectric and $k = \omega\sqrt
{\varepsilon_{0}\mu_{0}}$ is the wave number in vacuum, respectively. Using the Sommerfeld
identity \cite{sommerfeld, chewbook},
\begin{equation}
\frac{e^{ik_{1}|\mathbf{r}-\mathbf{r}^{\prime}|}}{|\mathbf{r}-\mathbf{r}%
^{\prime}|} = \frac{i}{2\pi} \int_{-\infty}^{+\infty}\int_{-\infty}^{+\infty}
dk_{x} dk_{y} \frac{e^{ik_{x}(x-x^{\prime})+ik_{y}(y-y^{\prime})+ik_{1z}%
|z-z^{\prime}|}}{k_{1z}},\label{sommerfeld_id}%
\end{equation}
where $k^{2}_{s} = k^{2}_{x}+k^{2}_{y}$ and $k_{1z} = \sqrt{k^{2}_{1}%
-k^{2}_{s}}$, the $E_{1z}^{P}$ can be written as
\begin{align}
E^{P}_{1z}(\mathbf{r})  & =\int_{-\infty}^{+\infty}\int_{-\infty}^{+\infty}
dk_{x} dk_{y} \left(  -\frac{\omega\mu_{0}\mu_{1} }{8\pi^{2}} \left(
\hat{z}\cdot\hat{\alpha}^{\prime}+\frac{1}{k_{1}^{2}} \partial_{z} \nabla\cdot
\hat{\alpha}^{\prime}\right)  \frac{e^{ik_{x}(x-x^{\prime})+ik_{y}%
(y-y^{\prime})+ik_{1z}|z-z^{\prime}|}}{k_{1z}}\right),\label{Ez}%
\end{align}
where $\hat{z}$ is a unit vector along the $z$-axis. The integrand in Eq. (\ref{Ez}) is the spectral component of electric field in
the $z$-direction, which is denoted by
\begin{align}
\tilde{E}^{P}_{1z} &  = \left(  \hat{z}\cdot\hat{\alpha}^{\prime}+\frac
{1}{k_{1}^{2}} \partial_{z} \nabla\cdot\hat{\alpha}^{\prime}\right)  \tilde
{g}_{1}^{P},\label{E_1z}%
\end{align}
where
\begin{equation}
\tilde{g}_{1}^{P} = -\frac{\omega\mu_{0}\mu_{1} }{8\pi^{2}} \frac
{e^{ik_{x}(x-x^{\prime})+ik_{y}(y-y^{\prime})+ik_{1z}|z-z^{\prime}|}}{k_{1z}}.
\end{equation}
A similar derivation yields the magnetic field as
\begin{equation}
\tilde{H}^{P}_{1z} = \frac{1}{i\omega\mu_{0}\mu_{1}}\hat{z} \cdot\nabla_{s}
\times\hat{\alpha}^{\prime}\tilde{g}_{1}^{P},\label{H_1z}%
\end{equation}
where $\nabla_{s} = (\partial_{x}, \partial_{y})$. From Maxwell's equations,
the transverse components $\tilde{\mathbf{E}}^{P}_{1s} = (\tilde{E}^{P}_{1x},
\tilde{E}^{P}_{1y})$ and $\tilde{\mathbf{H}}^{P}_{1s} = (\tilde{H}%
^{P}_{1x}, \tilde{H}^{P}_{1y})$ can be written using the $\tilde{E}_{1z}$
and $\tilde{H}_{1z}$ as
\begin{align}
\tilde{\mathbf{E}}^{P}_{1s}  & = \frac{1}{k_{s}^{2}} \left( \nabla_{s}
\partial_{z} \tilde{E}_{1z}^{P} - i\omega\mu_{0} \mu_{1} \hat{z} \times
\nabla_{s} \tilde{H}_{1z}^{P}\right) ,\label{E_1s}\\
\tilde{\mathbf{H}}^{P}_{1s}  & =\frac{1}{k_{s}^{2}} \left( \nabla_{s}
\partial_{z} \tilde{H}_{1z}^{P} +i\omega\varepsilon_{0} \varepsilon_{1}
\hat{z} \times\nabla_{s} \tilde{E}_{1z}^{P}\right) .\label{H_1s}%
\end{align}
These two relations reduce the problem to an one-dimensional problem in the spectral domain because only the $z$-component of electric and magnetic fields is required to completely determine the fields in the spectral domain. By substituting Eqs. (\ref{E_1z}) and (\ref{H_1z}) into
Eqs. (\ref{E_1s}) and (\ref{H_1s}), the electric field in the spectral domain
$\tilde{\mathbf{E}}^{P} = (\tilde{E}_{1x}^{P}, \tilde{E}_{1y}^{P}, \tilde
{E}_{1z}^{P})$ can be explicitly written in terms of the spectral Green's
function, that is,
\begin{align}
\left[
\begin{array}
[c]{c}%
\tilde{E}_{1x}^{P}\\
\tilde{E}_{1y}^{P}\\
\tilde{E}_{1z}^{P}%
\end{array}
\right]  = \tilde{\mathbf{G}}^{P} \boldsymbol{\alpha}^{\prime}= \left[
\begin{array}
[c]{ccc}%
\tilde{G}^{P}_{xx} & \tilde{G}^{P}_{xy} & \tilde{G}^{P}_{xz}\\
\tilde{G}^{P}_{yx} & \tilde{G}^{P}_{yy} & \tilde{G}^{P}_{yz}\\
\tilde{G}^{P}_{zx} & \tilde{G}^{P}_{zy} & \tilde{G}^{P}_{zz}%
\end{array}
\right]  \left[
\begin{array}
[c]{c}%
\alpha_{x}^{\prime}\\
\alpha_{y}^{\prime}\\
\alpha_{z}^{\prime}%
\end{array}
\right] ,
\end{align}
where
\begin{align}
\tilde{\mathbf{G}}^{P} = \left( \mathbf{I}+\frac{\nabla\nabla}{k_{1}^{2}%
}\right) \tilde{g}_{1}^{P}.\label{spectral_free}%
\end{align}
Finally, the electric field in the spatial domain in Eq. (\ref{primary_h}) can be
recovered by taking double integrals
\begin{align}
G^{P}_{ij} = \int_{-\infty}^{+\infty} \int_{-\infty}^{+\infty}\tilde{G}%
^{P}_{ij} dk_{x} dk_{y},~ i,j=x,y,z,
\end{align}
on each component of Eq. (\ref{spectral_free}). This will constitute the free-space dyadic Green's function in the spatial domain.
Note that this integral is well known as a Sommerfeld integral. The double
integral can be reduced to a single integral using cylindrical coordinate. The
resulting integral involves Bessel function or Hankel function depending on
convenience and is sometimes referred as the Hankel transform.
\begin{figure}[t]
\centering  \includegraphics[width=3in]{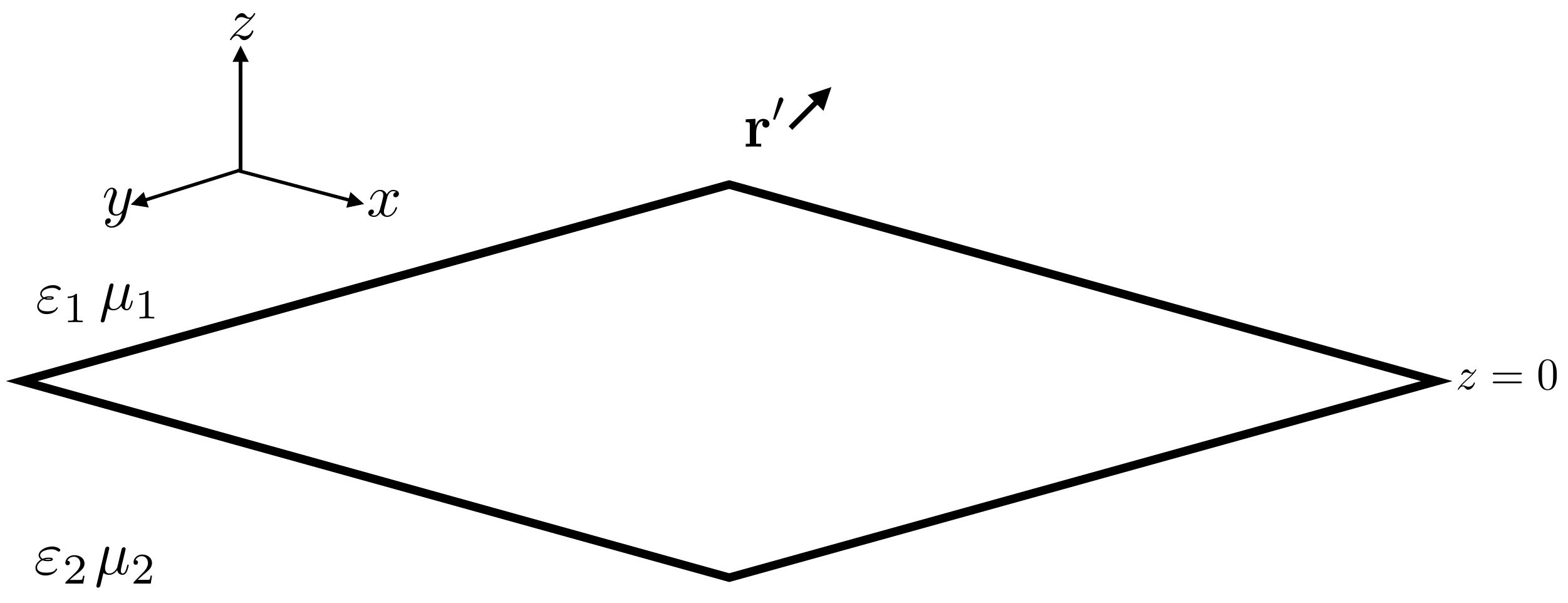}  \caption{A two-layer
structure. The free space is divided at $z = 0$ into the top and bottom layer.
A dipole is located at $\mathbf{r}^{\prime}= (x^{\prime},y^{\prime},z^{\prime
})$ and the top layer has $\varepsilon_{1}$ and $\mu_{1}$ and the bottom layer
has $\varepsilon_{2}$ and $\mu_{2}$. }%
\label{twolayer_fig}%
\end{figure}

\section{Green's function for a two-layer structure}

In this section, the free-space Green's function is modified with the
reflected and transmitted parts of the Green's function for a two-layer
structure depicted in Fig. \ref{twolayer_fig}. Overall the process of
computing the Green's function is the same as the free space. Due to symmetry,
the source is assumed to be in the first layer. First, in the spectral domain,
using Fresnel reflection and transmission coefficients, $z$-component of
reflected electric field in the first layer and the transmitted field in the
second layer are found. Then, all the transverse components in each layer are
derived using Eqs. (\ref{E_1s}) and (\ref{H_1s}). Now the spatial domain
Green's function can be found by taking Sommerfeld integral for each
component. Finally, in the first layer, primary field is added with reflected
field from the second layer to complete derivation.

\subsection{Fields in the spectral domain}

The $z$-component of electric and magnetic fields in the first layer are
\begin{align}
\tilde{E}_{1z} = \tilde{E}_{1z}^{P}+\tilde{E}_{1z}^{R},~~\tilde{H}_{1z} =
\tilde{H}_{1z}^{P}+\tilde{H}_{1z}^{R},%
\end{align}
where the superscript $P$ and $R$ denote the primary and reflected parts,
respectively. Similarly, The $z$-component of electric and magnetic fields in
the second layer are
\begin{align}
\tilde{E}_{2z} = \tilde{E}_{2z}^{T},~~\tilde{H}_{2z} = \tilde{H}_{2z}^{T},%
\end{align}
where the superscript $T$ denotes the transmitted part. The $\tilde{E}_{1z}^{P}$
and $\tilde{H}_{1z}^{P}$ are the primary fields given in the previous section.
Using the Fresnel reflection and transmission coefficients between the first and the second layer,
\begin{align}
R_{12}^{TM} = \frac{\varepsilon_{2} k_{1z} -\varepsilon_{1} k_{2z}
}{\varepsilon_{2} k_{1z} +\varepsilon_{1} k_{2z} }, \mbox{ } R_{12}^{TE} =
\frac{\mu_{2} k_{1z} -\mu_{1} k_{2z} }{\mu_{2} k_{1z} +\mu_{1} k_{2z} }, T_{12}^{TM} = \frac{2\varepsilon_{2} k_{2z}}{\varepsilon_{2}k_{1z}%
+\varepsilon_{1}k_{2z}}, \mbox{ } T_{12}^{TE} = \frac{2\mu_{2} k_{2z}}{\mu
_{2}k_{1z}+\mu_{1}k_{2z}},
\end{align}
the reflected and transmitted parts can be found as
\begin{align}
\tilde{E}_{1z}^{R}  & = \left(  \hat{z} \cdot\hat{\boldsymbol{\alpha}}%
^{\prime\prime}+\frac{1}{k_{1}^{2}} \partial_{z} \nabla\cdot\hat
{\boldsymbol{\alpha}}^{\prime\prime}\right)  \tilde{g}_{1, TM}^{R},~~
\tilde{E}_{2z}^{T} = \left( \hat{z} \cdot\hat{\boldsymbol{\alpha}}%
^{\prime\prime\prime}+\frac{1}{k_{2}^{2}} \partial_{z} \nabla\cdot
\hat{\boldsymbol{\alpha}}^{\prime\prime\prime}\right)  \tilde{g}_{2,TM}%
^{T},\label{e1z_r}\\
\tilde{H}_{1z}^{R}  & = -\frac{1}{i\omega\mu_{0}\mu_{1}}\hat{z} \cdot
\nabla_{s} \times\hat{\boldsymbol{\alpha}}^{\prime\prime}\tilde{g}_{1,TE}%
^{R},~~ \tilde{H}_{2z}^{T} = \frac{1}{i\omega\mu_{0}\mu_{2}} \frac{\mu_{1}%
}{\mu_{2}}\frac{k_{2z}}{k_{1z}}\hat{z} \cdot\nabla_{s} \times\hat
{\boldsymbol{\alpha}}^{\prime\prime\prime}\tilde{g}_{2,TE}^{T},\label{h1z_r}%
\end{align}
where
\begin{align}
\tilde{g}_{1, TM,TE}^{R}  & = -R_{12}^{TM, TE} \frac{\omega\mu_{0}\mu_{1}%
}{8\pi^{2}} \frac{e^{ik_{x}(x-x^{\prime})+ik_{y}(y-y^{\prime})+ik_{1z}%
(z+z^{\prime})}}{k_{1z}},\\
\tilde{g}_{2, TM,TE}^{T}  & = -T_{12}^{TM, TE} \frac{\omega\mu_{0}\mu_{2}%
}{8\pi^{2}} \frac{e^{ik_{x}(x-x^{\prime})+ik_{y}(y-y^{\prime})-ik_{2z}%
z+ik_{1z}z^{\prime}}}{k_{1z}},\\
\hat{\boldsymbol{\alpha}}^{\prime\prime} & = (-\alpha_{x}^{\prime},
-\alpha_{y}^{\prime}, \alpha_{z}), \hat{\boldsymbol{\alpha}}^{\prime
\prime\prime}= (\frac{k_{1z}}{k_{2z}}\alpha_{x}^{\prime}, \frac{k_{1z}}%
{k_{2z}}\alpha_{y}^{\prime}, \alpha_{z}).
\end{align}
The $\boldsymbol{\hat{\alpha}}^{\prime\prime}$ for the reflected part and
$\boldsymbol{\hat{\alpha}}^{\prime\prime\prime}$ for the transmitted part ensure the
boundary conditions between layers. The transverse component $\tilde
{\mathbf{E}}^{R}_{1s} = (\tilde{E}^{R}_{1x}, \tilde{E}^{R}_{1y})$ can be expressed using the $z$-components in Eqs. (\ref{e1z_r}) and
(\ref{h1z_r}) using Eq. (\ref{E_1s}) as before by simply replacing the
subindex with either $1$ or $2$ and superscript with either $R$ or $T$ as
\begin{align}
\tilde{\mathbf{E}}^{R}_{1s} = \frac{1}{k_{s}^{2}}\nabla_{s} \partial_{z}
\left( \hat{z} \cdot\hat{\boldsymbol{\alpha}}^{\prime\prime}+\frac{1}%
{k_{1}^{2}} \partial_{z} \nabla\cdot\hat{\boldsymbol{\alpha}}^{\prime\prime
}\right)  \tilde{g}_{1, TM}^{R} + \frac{1}{k_{s}^{2}}\hat{z}\times\nabla_{s}
\left( \hat{z} \cdot\nabla_{s}\times\hat{\boldsymbol{\alpha}}^{\prime\prime
}\right)  \tilde{g}_{1,TE}^{R}.
\end{align}
Then, $x$- and $y$-components can be explicitly written out as
\begin{align}
\tilde{E}^{R}_{1x}  &  = \frac{1}{k_{1}^{2}}\left( \frac{k_{1z}^{2}}{k_{s}^{2}
}\partial_{x}^{2}\tilde{g}_{1,TM}^{R}-\frac{k_{1}^{2}}{k_{s}^{2}}\partial
_{y}^{2} \tilde{g}_{1,TE}^{R} \right) \alpha_{x}^{\prime}+\frac{1}{k_{1}^{2}%
}\left( \frac{k_{1z}^{2}}{k_{s}^{2} }\partial_{x}\partial_{y} \tilde{g}%
_{1,TM}^{R}+\frac{k_{1}^{2}}{k_{s}^{2}}\partial_{x}\partial_{y} \tilde
{g}_{1,TE}^{R} \right) \alpha_{y}^{\prime}+\frac{1}{k_{1}^{2}}\partial_{x}
\partial_{z}\tilde{g}_{1,TM}^{R} \alpha_{z}^{\prime}\\
\tilde{E}^{R}_{1y}  &  = \frac{1}{k_{1}^{2}}\left( \frac{k_{1z}^{2}}{k_{s}^{2}
}\partial_{x} \partial_{y} \tilde{g}_{1,TM}^{R}+\frac{k_{1}^{2}}{k_{s}^{2}%
}\partial_{x} \partial_{y} \tilde{g}_{1,TE}^{R} \right) \alpha_{x}^{\prime
}+\frac{1}{k_{1}^{2}}\left( \frac{k_{1z}^{2}}{k_{s}^{2} }\partial_{y}^{2}
\tilde{g}_{1,TM}^{R}-\frac{k_{1}^{2}}{k_{s}^{2}}\partial_{x}^{2} \tilde
{g}_{1,TE}^{R} \right) \alpha_{y}^{\prime}+\frac{1}{k_{1}^{2}}\partial_{y}
\partial_{z}\tilde{g}_{1,TM}^{R} \alpha_{z}^{\prime}%
\end{align}
By listing all the components of the electric field in the spectral domain, the Green's function in the first layer can be found as
\begin{align}
\left[
\begin{array}
[c]{c}%
\tilde{E}_{1x}\\
\tilde{E}_{1y}\\
\tilde{E}_{1z}%
\end{array}
\right]   & = \left[
\begin{array}
[c]{c}%
\tilde{E}_{1x}^{P}+\tilde{E}_{1x}^{R}\\
\tilde{E}_{1y}^{P}+\tilde{E}_{1y}^{R}\\
\tilde{E}_{1z}^{P}+\tilde{E}_{1z}^{R}%
\end{array}
\right]  =\left( \tilde{\mathbf{G}}^{P}-\frac{1}{8\pi^{2}\omega\varepsilon
_{0}\varepsilon_{1}}\tilde{\mathbf{G}}_{1}^{R}\right)
\boldsymbol{\alpha}^{\prime}\nonumber\\
& = \left( \left[
\begin{array}
[c]{ccc}%
\tilde{G}^{P}_{xx} & \tilde{G}^{P}_{xy} & \tilde{G}^{P}_{xz}\\
\tilde{G}^{P}_{yx} & \tilde{G}^{P}_{yy} & \tilde{G}^{P}_{yz}\\
\tilde{G}^{P}_{zx} & \tilde{G}^{P}_{zy} & \tilde{G}^{P}_{zz}%
\end{array}
\right]  - \frac{1}{8\pi^{2}\omega\varepsilon_{0}\varepsilon_{1}}\left[
\begin{array}
[c]{ccc}%
\tilde{G}^{R}_{1xx} & \tilde{G}^{R}_{1xy} & \tilde{G}^{R}_{1xz}\\
\tilde{G}^{R}_{1yx} & \tilde{G}^{R}_{1yy} & \tilde{G}^{R}_{1yz}\\
\tilde{G}^{R}_{1zx} & \tilde{G}^{R}_{1zy} & \tilde{G}^{R}_{1zz}%
\end{array}
\right]  \right) \left[
\begin{array}
[c]{c}%
\alpha_{x}^{\prime}\\
\alpha_{y}^{\prime}\\
\alpha_{z}^{\prime}%
\end{array}
\right] ,
\end{align}
where $\tilde{\mathbf{G}}^{P}$ is the same as Eq. (\ref{spectral_free}) and
all the components of $\tilde{\mathbf{G}}^{R}_{1}$ are given by
\begin{align}
\tilde{G}^{R}_{1xx} & = \left(  \partial_{x}^{2} \frac{k_{1z}}{k_{s}^{2}%
}R_{12}^{TM}-\partial_{y}^{2} \frac{k_{1}^{2}}{k_{s}^{2} k_{1z}}R_{12}^{TE}
\right) e^{ik_{x}(x-x^{\prime})+ik_{y}(y-y^{\prime})+ik_{1z}(z+z^{\prime}%
)},\label{1xx}\\
\tilde{G}^{R}_{1yy} & = \left(  \partial_{y}^{2} \frac{k_{1z}}{k_{s}^{2}%
}R_{12}^{TM}-\partial_{x}^{2} \frac{k_{1}^{2}}{k_{s}^{2} k_{1z}}R_{12}^{TE}
\right) e^{ik_{x}(x-x^{\prime})+ik_{y}(y-y^{\prime})+ik_{1z}(z+z^{\prime}%
)},\label{1yy}\\
\tilde{G}_{1zz}^{R}  &  = \left(  \frac{k_{s}^{2}}{k_{1z}} R_{12}^{TM}\right)
e^{ik_{x}(x-x^{\prime})+ik_{y}(y-y^{\prime})+ik_{1z}(z+z^{\prime})}
,\label{1zz}\\
\tilde{G}_{1xy}^{R} & =\tilde{G}_{1yx}^{R} = \partial_{x}\partial_{y} \left(
\frac{k_{1z}}{k_{s}^{2} }R_{12}^{TM}+\frac{k_{1}^{2}}{k_{s}^{2}k_{1z}}%
R_{12}^{TE}\right)  e^{ik_{x}(x-x^{\prime})+ik_{y}(y-y^{\prime})+ik_{1z}%
(z+z^{\prime})},\label{1xy}\\
\tilde{G}_{1xz}^{R}  & = -\tilde{G}_{1zx}^{R}= \partial_{x} \partial_{z}
\left(  \frac{R_{12}^{TM}}{k_{1z}} \right)  e^{ik_{x}(x-x^{\prime}%
)+ik_{y}(y-y^{\prime})+ik_{1z}(z+z^{\prime})} ,\label{1xz}\\
\tilde{G}_{1yz}^{R}  & = -\tilde{G}_{1zy}^{R} = \partial_{y} \partial_{z}
\left(  \frac{R_{12}^{TM}}{k_{1z}} \right)  e^{ik_{x}(x-x^{\prime}%
)+ik_{y}(y-y^{\prime})+ik_{1z}(z+z^{\prime})} .\label{1yz}%
\end{align}
In the second layer, the transverse component  $\tilde{\mathbf{E}}^{T}_{2s} = (\tilde{E}^{T}_{2x}, \tilde{E}^{T}_{2y})$ is
\begin{align}
\tilde{\mathbf{E}}^{T}_{2s} = \frac{1}{k_{s}^{2}}\nabla_{s} \partial_{z}
\left( \hat{z} \cdot\hat{\boldsymbol{\alpha}}^{\prime\prime\prime}+\frac
{1}{k_{2}^{2}} \partial_{z} \nabla\cdot\hat{\boldsymbol{\alpha}}^{\prime
\prime\prime}\right)  \tilde{g}_{2, TM}^{T} - \frac{1}{k_{s}^{2}}\frac{\mu
_{1}}{\mu_{2}}\frac{k_{2z}}{k_{1z}}\hat{z}\times\nabla_{s} \left( \hat{z}
\cdot\nabla_{s}\times\hat{\boldsymbol{\alpha}}^{\prime\prime\prime}\right)
\tilde{g}_{2,TE}^{T}.
\end{align}
Then, $x$- and $y$-components can be explicitly written out as
\begin{align}
\tilde{E}_{2x}^{T}  &  = \frac{1}{k_{2}^{2}} \left( -\partial_{x}^{2}
\frac{k_{1z}k_{2z}}{k_{s}^{2}} \tilde{g}_{2, TM}^{T} -\partial_{y}^{2}
\frac{k_{2}^{2}}{k_{s}^{2}}\frac{\mu_{1}}{\mu_{2}} \tilde{g}_{2, TE}^{T}
\right)  \alpha_{x}^{\prime}+\frac{1}{k_{2}^{2}} \left(  -\partial_{x}
\partial_{y} \frac{k_{1z}k_{2z}}{k_{s}^{2}}\tilde{g}_{2, TM}^{T} +\partial_{x}
\partial_{y} \frac{k_{2}^{2}}{k_{s}^{2}}\frac{\mu_{1}}{\mu_{2}} \tilde
{g}_{2,TE}^{T} \right)  \alpha_{y}^{\prime}\nonumber\\
& +\frac{1}{k_{2}^{2}}\partial_{x} \partial_{z} \tilde{g}_{2, TM}^{T}
\alpha_{z}^{\prime},
\end{align}
\begin{align}
\tilde{E}_{2y}^{T}  &  = \frac{1}{k_{2}^{2}} \left(  -\partial_{x}
\partial_{y} \frac{k_{1z}k_{2z}}{k_{s}^{2}}\tilde{g}_{2, TM}^{T} +\partial_{x}
\partial_{y} \frac{k_{2}^{2}}{k_{s}^{2}}\frac{\mu_{1}}{\mu_{2}} \tilde
{g}_{2,TE}^{T} \right)  \alpha_{x}^{\prime}+\frac{1}{k_{2}^{2}} \left(
-\partial_{y}^{2} \frac{k_{1z}k_{2z}}{k_{s}^{2}} \tilde{g}_{2, TM}^{T}
-\partial_{x}^{2} \frac{k_{2}^{2}}{k_{s}^{2}}\frac{\mu_{1}}{\mu_{2}} \tilde
{g}_{2, TE}^{T} \right)  \alpha_{y}^{\prime}\nonumber\\
& +\frac{1}{k_{2}^{2}}\partial_{y} \partial_{z} \tilde{g}_{2, TM}^{T}
\alpha_{z}^{\prime}.
\end{align}
Thus, the spectral Green's function in the second layer is
\begin{align}
\left[
\begin{array}
[c]{c}%
\tilde{E}_{2x}\\
\tilde{E}_{2y}\\
\tilde{E}_{2z}%
\end{array}
\right]   & = \left[
\begin{array}
[c]{c}%
\tilde{E}_{2x}^{T}\\
\tilde{E}_{2y}^{T}\\
\tilde{E}_{2z}^{T}%
\end{array}
\right]  =-\frac{1}{8\pi^{2} \omega\varepsilon_{0} \varepsilon_{2}}%
\tilde{\mathbf{G}}_{2}^{T} \hat{\boldsymbol{\alpha}}^{\prime}= -\frac{1}%
{8\pi^{2} \omega\varepsilon_{0} \varepsilon_{2}}\left[
\begin{array}
[c]{ccc}%
\tilde{G}^{T}_{2xx} & \tilde{G}^{T}_{2xy} & \tilde{G}^{T}_{2xz}\\
\tilde{G}^{T}_{2yx} & \tilde{G}^{T}_{2yy} & \tilde{G}^{T}_{2yz}\\
\tilde{G}^{T}_{2zx} & \tilde{G}^{T}_{2zy} & \tilde{G}^{T}_{2zz}%
\end{array}
\right]  \left[
\begin{array}
[c]{c}%
\alpha_{x}^{\prime}\\
\alpha_{y}^{\prime}\\
\alpha_{z}^{\prime}%
\end{array}
\right] ,
\end{align}
where
\begin{align}
\tilde{G}_{2xx}^{T}  &  = \left( -\partial_{x}^{2} \frac{k_{2z}}{k_{s}^{2}}
T_{12}^{TM} -\partial_{y}^{2} \frac{k_{2}^{2}}{k_{1z}k_{s}^{2}}\frac{\mu_{1}%
}{\mu_{2}} T_{12}^{TE}\right) e^{ik_{x}(x-x^{\prime})+ik_{y}(y-y^{\prime
})-ik_{2z}z+ik_{1z}z^{\prime}},\label{2xx}\\
\tilde{G}_{2yy}^{T}  & =\left( -\partial_{y}^{2} \frac{k_{2z}}{k_{s}^{2}}
T_{12}^{TM} -\partial_{x}^{2} \frac{k_{2}^{2}}{k_{1z}k_{s}^{2}}\frac{\mu_{1}%
}{\mu_{2}} T_{12}^{TE} \right)  e^{ik_{x}(x-x^{\prime})+ik_{y}(y-y^{\prime
})-ik_{2z}z+ik_{1z}z^{\prime}},\\
\tilde{G}_{2zz}^{T}  & =\left(  \frac{k_{s}^{2}}{k_{1z}}T_{12}^{TM} \right)
e^{ik_{x}(x-x^{\prime})+ik_{y}(y-y^{\prime})-ik_{2z}z+ik_{1z}z^{\prime}},\\
\tilde{G}_{2xy}^{T}  &  = \tilde{G}_{2, yx}^{T} = \left(  -\partial_{x}
\partial_{y} \frac{k_{2z}}{k_{s}^{2}} T_{12}^{TM} +\partial_{x} \partial_{y}
\frac{k_{2}^{2}}{k_{1z}k_{s}^{2}}\frac{\mu_{1}}{\mu_{2}} T_{12}^{TE} \right)
e^{ik_{x}(x-x^{\prime})+ik_{y}(y-y^{\prime})-ik_{2z}z+ik_{1z}z^{\prime}},\\
\tilde{G}_{2xz}^{T}  & = \partial_{x} \partial_{z} \frac{T_{12}^{TM}}{k_{1z}}
e^{ik_{x}(x-x^{\prime})+ik_{y}(y-y^{\prime})-ik_{2z}z+ik_{1z}z^{\prime}},
\tilde{G}_{2yz}^{T} = \partial_{y} \partial_{z} \frac{T_{12}^{TM}}{k_{1z}}
e^{ik_{x}(x-x^{\prime})+ik_{y}(y-y^{\prime})-ik_{2z}z+ik_{1z}z^{\prime}},\\
\tilde{G}_{2zx}^{T}  & = \partial_{z} \partial_{x} \frac{T_{12}^{TM}}{k_{2z}}
e^{ik_{x}(x-x^{\prime})+ik_{y}(y-y^{\prime})-ik_{2z}z+ik_{1z}z^{\prime}%
},\tilde{G}_{2zy}^{T} = \partial_{z} \partial_{y} \frac{T_{12}^{TM}}{k_{2z}}
e^{ik_{x}(x-x^{\prime})+ik_{y}(y-y^{\prime})-ik_{2z}z+ik_{1z}z^{\prime}}.
\end{align}

\subsection{Fields in the spatial domain}

In this subsection, Sommerfeld integrals/Hankel transforms are used to convert
the spectral Green's function found in the previous subsection to the spatial
domain Green's function. Several useful Bessel identities are listed in \ref{besselidentity}
and used in the derivation. \newline

\noindent$\bullet$ In the first layer, the first component $G_{1xx}^{R}$ can be found by taking double integral as
follows
\begin{align}
G_{1xx}^{R}  & = \int_{-\infty}^{+\infty} \int_{-\infty}^{+\infty} \tilde
{G}_{1xx}^{R} dk_{x} dk_{y}\nonumber\\
& =\int_{-\infty}^{+\infty} \int_{-\infty}^{+\infty} \left(  \partial_{x}^{2}
\frac{k_{1z}}{k_{s}^{2}}R_{12}^{TM}-\partial_{y}^{2} \frac{k_{1}^{2}}%
{k_{s}^{2} k_{1z}}R_{12}^{TE} \right) e^{ik_{x}(x-x^{\prime})+ik_{y}%
(y-y^{\prime})+ik_{1z}(z+z^{\prime})}dk_{x} dk_{y}\nonumber\\
&  = \mbox{\textcircled{1}}-\mbox{\textcircled{2}} ,
\end{align}
where
\begin{align}
\mbox{\textcircled{1}}  & = \int_{-\infty}^{+\infty} \int_{-\infty}^{+\infty}
\partial_{x}^{2} \frac{k_{1z}}{k_{s}^{2}}R_{12}^{TM}e^{ik_{x}(x-x^{\prime
})+ik_{y}(y-y^{\prime})+ik_{1z}(z+z^{\prime})} dk_{x} dk_{y},\\
\mbox{\textcircled{2}}  & = \int_{-\infty}^{+\infty} \int_{-\infty}^{+\infty}
\partial_{y}^{2} \frac{k_{1}^{2}}{k_{s}^{2} k_{1z}}R_{12}^{TE}e^{ik_{x}%
(x-x^{\prime})+ik_{y}(y-y^{\prime})+ik_{1z}(z+z^{\prime})} dk_{x} dk_{y}.
\end{align}
For all double integrals throughout the paper, the cylindrical coordinate
transform
\begin{align}
k_{x}  & = k_{s}\cos{\phi}, k_{y} = k_{s}\sin{\phi}, (x-x^{\prime}) = \rho
\cos{\theta}, (y-y^{\prime}) = \rho\sin{\theta},
\end{align}
is used to reduce double integrals into single integrals. Now the integral
\textcircled{1} is simplified as
\begin{align}
\mbox{\textcircled{1}}  & = \int_{-\infty}^{+\infty} \int_{-\infty}^{+\infty}
\partial_{x}^{2} \frac{k_{1z}}{k_{s}^{2}}R_{12}^{TM}e^{ik_{x}(x-x^{\prime
})+ik_{y}(y-y^{\prime})+ik_{1z}(z+z^{\prime})} dk_{x} dk_{y},\nonumber\\
& =-\int_{0}^{\infty}k_{s} k_{1z}R_{12}^{TM}e^{ik_{1z}(z+z^{\prime})}
\int_{0}^{2\pi} e^{ik_{s}\rho\cos{(\phi-\theta)}} \cos^{2}{\phi} d\phi dk_{s}
\nonumber\\
& =-\int_{0}^{\infty}k_{s} k_{1z}R_{12}^{TM}e^{ik_{1z}(z+z^{\prime})} \left(
\pi J_{0}(k_{s}\rho) - \pi J_{2}(k_{s}\rho) \cos{2\theta}\right)
dk_{s}\nonumber\\
& =-\frac{1}{2}2\pi\int_{0}^{\infty}k_{s} \tilde{g}_{1,1}^{R} J_{0}(k_{s}%
\rho)e^{ik_{1z}(z+z^{\prime})}dk_{s}+\pi(1-2\sin^{2}{\theta})\int_{0}^{\infty}
k_{s}^{3} \frac{k_{1z}}{k_{s}^{2}} R_{12}^{TM} J_{2}(k_{s}\rho)e^{ik_{1z}%
(z+z^{\prime})}dk_{s}\nonumber\\
& =-\frac{1}{2} g_{1,1}^{R}+\pi\left( 1-2\frac{(y-y^{\prime})^{2}}{\rho^{2}%
}\right) \int_{0}^{\infty} k_{s}^{3} \tilde{g}_{1,2}^{R} J_{2}(k_{s}%
\rho)e^{ik_{1z}(z+z^{\prime})}dk_{s}\nonumber\\
& =-\frac{1}{2} g_{1,1}^{R}+ \left( \frac{1}{2}\rho^{2}-(y-y^{\prime}%
)^{2}\right)  2\pi\int_{0}^{\infty} k^{3}_{s} \tilde{g}_{1,2}^{R} \frac
{J_{2}(k_{s}\rho)}{\rho^{2}}e^{ik_{1z}(z+z^{\prime})}dk_{s}\nonumber\\
& =-\frac{1}{2} g_{1,1}^{R}+ \left( \frac{1}{2}\rho^{2}-(y-y^{\prime}%
)^{2}\right)  g_{1,2}^{R},
\end{align}
where
\begin{align}
g_{1,1}^{R}  & = 2\pi\int_{0}^{\infty} k_{s} \tilde{g}_{1,1}^{R} J_{0}%
(k_{s}\rho) e^{ik_{1z}(z+z^{\prime})} dk_{s}, g_{1,2}^{R} = 2\pi\int
_{0}^{\infty} k_{s}^{3} \tilde{g}_{1,2}^{R} \frac{J_{2}(k_{s}\rho)}{\rho^{2}}
e^{ik_{1z}(z+z^{\prime})} dk_{s},\nonumber\\
\tilde{g}_{1,1}^{R}  & = k_{1z}R_{12}^{TM}, \tilde{g}_{1,2}^{R} = \frac
{k_{1z}R_{12}^{TM}}{k_{s}^{2}}.
\end{align}
Similar derivation yields the integral \textcircled{2} as follows
\begin{align}
\mbox{\textcircled{2}}  & = \int_{-\infty}^{+\infty} \int_{-\infty}^{+\infty}
\partial_{y}^{2} \frac{k_{1}^{2}}{k_{s}^{2} k_{1z}}R_{12}^{TE}e^{ik_{x}%
(x-x^{\prime})+ik_{y}(y-y^{\prime})+ik_{1z}(z+z^{\prime})} dk_{x}
dk_{y}\nonumber\\
& =-\frac{1}{2}g_{1,3}^{R}- \left( \frac{1}{2}\rho^{2}-(y-y^{\prime})^{2}
\right) g_{1,4}^{R}.
\end{align}
where
\begin{align}
g_{1,3}^{R}  & = 2\pi\int_{0}^{\infty} k_{s} \tilde{g}_{1,3}^{R} J_{0}%
(k_{s}\rho) e^{ik_{1z}(z+z^{\prime})} dk_{s}, g_{1,4}^{R} = 2\pi\int
_{0}^{\infty} k_{s}^{3} \tilde{g}_{1,3}^{R} \frac{J_{2}(k_{s}\rho)}{\rho^{2}}
e^{ik_{1z}(z+z^{\prime})} dk_{s}, \nonumber\\
\tilde{g}_{1,3}^{R}  & = \frac{k_{1}^{2}}{k_{1z}}R_{12}^{TE},\tilde{g}%
_{1,4}^{R} = \frac{k_{1}^{2}}{k_{s}^{2}k_{1z}}R_{12}^{TE}.%
\end{align}
Therefore
\begin{align}
G_{xx}^{R}  & = -\frac{1}{2} g_{1,1}^{R}+ \left( \frac{1}{2}\rho
^{2}-(y-y^{\prime})^{2}\right)  g_{1,2}^{R} +\frac{1}{2}g_{1,3}^{R}+\left(
\frac{1}{2}\rho^{2}-(y-y^{\prime})^{2} \right) g_{1,4}^{R}\nonumber\\
& =-\frac{1}{2} \left( g_{1,1}^{R}-g_{1,3}^{R}\right) + \left( \frac{1}{2}%
\rho^{2}-(y-y^{\prime})^{2}\right)  \left( g_{1,2}^{R} +g_{1,4}^{R}\right)
=-\frac{1}{2}g_{1,5}^{R}+ \left( \frac{1}{2}\rho^{2}-(y-y^{\prime}%
)^{2}\right)  g_{1,6}^{R},
\end{align}
where
\begin{align}
g_{1,5}^{R}  & = 2\pi\int_{0}^{\infty} k_{s} \tilde{g}_{1,5}^{R} J_{0}%
(k_{s}\rho) e^{ik_{1z}(z+z^{\prime})} dk_{s}, g_{1,6}^{R} = 2\pi\int
_{0}^{\infty} k_{s}^{3} \tilde{g}_{1,6}^{R} \frac{J_{2}(k_{s}\rho)}{\rho^{2}}
e^{ik_{1z}(z+z^{\prime})} dk_{s},\nonumber\\
\tilde{g}_{1,5}^{R}  &  = k_{1z}R_{12}^{TM} - \frac{k_{1}^{2}}{k_{1z}}%
R_{12}^{TE}, \tilde{g}_{1,6}^{R} =\frac{k_{1z}R_{12}^{TM}}{k_{s}^{2}}%
+\frac{k_{1}^{2}}{k_{s}^{2}k_{1z}}R_{12}^{TE} .\label{g16}%
\end{align}
Thus, the $G_{xx}^{R}$ can be computed with only two Sommerfeld integrals.

Absolutely, the same derivation applies to $G_{1yy}^{R}$, that is,
\begin{align}
G_{1yy}^{R}  & = \int_{-\infty}^{+\infty} \int_{-\infty}^{+\infty} \tilde
{G}_{1yy}^{R} dk_{x} dk_{y}\nonumber\\
& =\int_{-\infty}^{+\infty} \int_{-\infty}^{+\infty} \left(  \partial_{y}^{2}
\frac{k_{1z}}{k_{s}^{2}}R_{12}^{TM}-\partial_{x}^{2} \frac{k_{1}^{2}}%
{k_{s}^{2} k_{1z}}R_{12}^{TE} \right) e^{ik_{x}(x-x^{\prime})+ik_{y}%
(y-y^{\prime})+ik_{1z}(z+z^{\prime})}dk_{x} dk_{y}\nonumber\\
&  = -\frac{1}{2}g_{1,5}^{R} - \left( \frac{1}{2}\rho^{2} -(y-y^{\prime}%
)^{2}\right)  g_{1,6}^{R}.
\end{align}
The derivation of $G_{1zz}^{R}$ is straightforward as there is no derivative in $\tilde{G}_{1zz}^{R}$.
\begin{align}
G_{1zz}^{R}  & = \int_{-\infty}^{+\infty} \int_{-\infty}^{+\infty} \left(
\frac{k_{s}^{2}}{k_{1z}} R_{12}^{TM}\right) e^{ik_{x}(x-x^{\prime}%
)+ik_{y}(y-y^{\prime})+ik_{1z}(z+z^{\prime})} dk_{x} dk_{y}\nonumber\\
&  = \int_{0}^{\infty} k_{s} \tilde{g}_{1,7}^{R} e^{ik_{1z}(z+z^{\prime})}
\int_{0}^{2\pi} e^{ik_{s}\rho\cos{(\phi-\theta)}} d\phi dk_{s}\nonumber\\
&  = 2\pi\int_{0}^{\infty} k_{s} \tilde{g}_{1,7}^{R} J_{0}(k_{s}\rho)
e^{ik_{1z}(z+z^{\prime})} dk_{s} = g_{1,7}^{R},
\end{align}
where
\begin{align}
g_{1,7}^{R}  & = 2\pi\int_{0}^{\infty} k_{s} \tilde{g}_{1,7}^{R} J_{0}%
(k_{s}\rho) e^{ik_{1z}(z+z^{\prime})} dk_{s}, \tilde{g}_{1,7}^{R} =
\frac{k_{s}^{2}}{k_{1z}} R_{12}^{TM}.
\end{align}
The $G_{1xy}^{R} = G_{1yx}^{R}$ can be derived using $\tilde{g}_{1,6}^{R}$
that is already defined in Eq. (\ref{g16}) as
\begin{align}
G_{1xy}^{R}  & = G_{1yx}^{R} = \int_{-\infty}^{+\infty}\int_{-\infty}%
^{+\infty} \partial_{x}\partial_{y} \left( \frac{k_{1z}}{k_{s}^{2} }%
R_{12}^{TM}+\frac{k_{1}^{2}}{k_{s}^{2}k_{1z}}R_{12}^{TE}\right)
e^{ik_{x}(x-x^{\prime})+ik_{y}(y-y^{\prime})+ik_{1z}(z+z^{\prime})}dk_{x}
dk_{y}\nonumber\\
& = \int_{0}^{\infty}\int_{0}^{2\pi} -k_{s}^{2} \cos{\phi}\sin{\phi}
\tilde{g}_{1,6}^{R} e^{ik_{s}\rho\cos{(\phi-\theta)}}e^{ik_{1z}(z+z^{\prime}%
)}k_{s} dk_{s} d\phi\nonumber\\
& = -\frac{1}{2}\int_{0}^{\infty} k_{s}^{3} \tilde{g}_{1,6}^{R}%
e^{ik_{1z}(z+z^{\prime})} \int_{0}^{2\pi} e^{ik_{s}\rho\cos{(\phi-\theta)}%
}\sin{2\phi} d\phi dk_{s}\nonumber\\
& = -\frac{1}{2}\int_{0}^{\infty} k_{s}^{3}\tilde{g}_{1,6}^{R}e^{ik_{1z}%
(z+z^{\prime})} \left( -2\pi J_{2}(k_{s}\rho)\sin{2\theta} \right)
dk_{s}\nonumber\\
& = \sin{\theta}\cos{\theta} 2\pi\int_{0}^{\infty} k_{s}^{3} \tilde{g}%
_{1,6}^{R} J_{2}(k_{s}\rho)e^{ik_{1z}(z+z^{\prime})} dk_{s}\nonumber\\
& = (x-x^{\prime})(y-y^{\prime}) 2\pi\int_{0}^{\infty} k_{s}^{3} \tilde
{g}_{1,6}^{R} \frac{J_{2}(k_{s}\rho)}{\rho^{2}}e^{ik_{1z}(z+z^{\prime})}
dk_{s} = (x-x^{\prime})(y-y^{\prime}) g_{1,6}^{R}.%
\end{align}
The $G_{1xz}^{R}$ and $G_{1zx}^{R}$ can be derived at the same time as
\begin{align}
G_{1xz}^{R}  &  = -G_{1zx}^{R} = \int_{-\infty}^{+\infty} \int_{-\infty
}^{+\infty} \partial_{x} \partial_{z} \frac{R_{12}^{TM}}{k_{1z}}
e^{ik_{x}(x-x^{\prime})+ik_{y}(y-y^{\prime})+ik_{1z}(z+z^{\prime})} dk_{x}
dk_{y}\nonumber\\
&  = -\int_{0}^{\infty}k^{2}_{s} R_{12}^{TM}e^{ik_{1z}(z+z^{\prime})}
\int_{0}^{2\pi} e^{ik_{s}\rho\cos{(\phi-\theta)}} \cos{\phi} d\phi
dk_{s}\nonumber\\
&  = -2\pi i \cos{\theta} \int_{0}^{\infty}k^{2}_{s} R_{12}^{TM}J_{1}%
(k_{s}\rho) e^{ik_{1z}(z+z^{\prime})} dk_{s}\nonumber\\
&  = -2\pi i (x-x^{\prime}) \int_{0}^{\infty}k^{2}_{s} \tilde{g}_{1,8}^{R}
\frac{J_{1}(k_{s}\rho)}{\rho} e^{ik_{1z}(z+z^{\prime})} dk_{s} =
-i(x-x^{\prime})g_{1,8}^{R},
\end{align}
where
\begin{align}
g_{1,8}^{R} =2\pi\int_{0}^{\infty} k_{s}^{2} \tilde{g}_{1,8}^{R} \frac
{J_{1}(k_{s}\rho)}{\rho} e^{ik_{1z}(z+z^{\prime})}dk_{s}, \tilde{g}_{1,8}^{R}
= R_{12}^{TM}.
\end{align}
Finally, the $G_{1yz}^{R}$ and $G_{1zy}^{R}$ can be derived using symmetry as
\begin{align}
G_{1yz}^{R}  & = -G_{1zy}^{R} = \int_{-\infty}^{+\infty}\int_{-\infty}^{+\infty}
\partial_{y} \partial_{z} \frac{R_{12}^{TM}}{k_{1z}} e^{ik_{x}(x-x^{\prime
})+ik_{y}(y-y^{\prime})+ik_{1z}(z+z^{\prime})}dk_{x} dk_{y} =-i(y-y^{\prime})g_{1,8}^{R}.
\end{align}
\newline
\noindent$\bullet$ In the second layer, the derivation of transmitted part $\mathbf{G}_{2}^{T}$ is absolutely similar to the reflected part. Therefore, most of derivations are omitted unless there are notable differences.
\begin{align}
G_{2xx}^{T}  &  = \int_{-\infty}^{+\infty}\int_{-\infty}^{+\infty}\left(
-\partial_{x}^{2} \frac{k_{2z}}{k_{s}^{2}} T_{12}^{TM} -\partial_{y}^{2}
\frac{k_{2}^{2}}{k_{1z}k_{s}^{2}}\frac{\mu_{1}}{\mu_{2}} T_{12}^{TE}\right)
e^{ik_{x}(x-x^{\prime})+ik_{y}(y-y^{\prime})-ik_{2z}z+ik_{1z}z^{\prime}}
dk_{x} dk_{y}\\
&  = \frac{1}{2}g_{2,5}^{T} - (\frac{1}{2}\rho^{2}-(y-y^{\prime})^{2})
g_{2,6}^{T},
\end{align}
where
\begin{align}
g_{2,5}^{T}  & = 2\pi\int_{0}^{\infty} k_{s} \tilde{g}_{2,5}^{T}J_{0}%
(k_{s}\rho)e^{-ik_{2z}+ik_{1}z^{\prime}}dk_{s}, g_{2,6}^{T} = 2\pi\int
_{0}^{\infty} k_{s}^{3} \tilde{g}_{2,6}^{T} \frac{J_{2}(k_{s}\rho)}{\rho^{2}%
}e^{-ik_{2z}+ik_{1}z^{\prime}}dk_{s},\\
\tilde{g}_{2,5}^{T}  & = k_{2z} T_{12}^{TM}+\frac{k_{2}^{2}}{k_{1z}}\frac
{\mu_{1}}{\mu_{2}} T_{12}^{TE}, \tilde{g}_{2,6}^{T} = \frac{k_{2z} T_{12}%
^{TM}}{k_{s}^{2}}-\frac{k_{2}^{2}}{k_{1z}k_{s}^{2}}\frac{\mu_{1}}{\mu_{2}}
T_{12}^{TE}.
\end{align}
Similarly,
\begin{align}
G_{2yy}^{T}  &  = \int_{-\infty}^{+\infty}\int_{-\infty}^{+\infty}\left(
-\partial_{y}^{2} \frac{k_{2z}}{k_{s}^{2}} T_{12}^{TM} -\partial_{x}^{2}
\frac{k_{2}^{2}}{k_{1z}k_{s}^{2}}\frac{\mu_{1}}{\mu_{2}} T_{12}^{TE} \right)
e^{ik_{x}(x-x^{\prime})+ik_{y}(y-y^{\prime})-ik_{2z}z+ik_{1z}z^{\prime}%
}\nonumber\\
&  = \frac{1}{2} g_{2,5}^{T} + (\frac{1}{2}\rho^{2}-(y-y^{\prime})^{2})
g_{2,6}^{T}.
\end{align}
The $G_{2zz}^{T}$ can be obtained by
\begin{align}
G_{2zz}^{T}  & = \int_{-\infty}^{+\infty}\int_{-\infty}^{+\infty} \left(
\frac{k_{s}^{2}}{k_{1z}}T_{12}^{TM} \right)  e^{ik_{x}(x-x^{\prime}%
)+ik_{y}(y-y^{\prime})-ik_{2z}z+ik_{1z}z^{\prime}} dk_{x} dk_{y}\nonumber\\
& = 2\pi\int_{0}^{\infty} k_{s} \tilde{g}_{2,7}^{T} J_{0}(k_{s}\rho)
e^{-ik_{2z}z+ik_{1z}z^{\prime}} dk_{s} = g_{2,7}^{T},
\end{align}
where
\begin{align}
g_{2,7}^{T}  & = 2\pi\int_{0}^{\infty} k_{s} \tilde{g}_{2,7}^{T}J_{0}%
(k_{s}\rho)e^{-ik_{2z}+ik_{1}z^{\prime}}dk_{s}, \tilde{g}_{2,7}^{T} =
\frac{k_{s}^{2}}{k_{1z}} T_{12}^{TM}.
\end{align}
Again, the $G_{2xy}^{T} = G_{2yx}^{T} $ can be written using $\tilde{g}_{2,6}^{T}$ as
\begin{align}
G_{2xy}^{T}  & = G_{2yx}^{T} = \int_{-\infty}^{+\infty}\int_{-\infty}%
^{+\infty} -\partial_{x} \partial_{y} \left(  \frac{k_{2z}}{k_{s}^{2}}
T_{12}^{TM} - \frac{k_{2}^{2}}{k_{1z}k_{s}^{2}}\frac{\mu_{1}}{\mu_{2}}
T_{12}^{TE} \right) e^{ik_{x}(x-x^{\prime})+ik_{y}(y-y^{\prime})-ik_{2z}%
z+ik_{1z}z^{\prime}} dk_{x} dk_{y}\nonumber\\
& = \int_{-\infty}^{+\infty}\int_{-\infty}^{+\infty} k_{x} k_{y} \tilde
{g}_{2,6}^{T} e^{ik_{x}(x-x^{\prime})+ik_{y}(y-y^{\prime})-ik_{2z}%
z+ik_{1z}z^{\prime}} dk_{x} dk_{y} = -(x-x^{\prime})(y-y^{\prime})g_{2,6}^{T}.%
\end{align}
The $G_{2xz}^{T}$ and $G_{2yz}^{T}$ can be found at the same time using their symmetry
\begin{align}
G_{2xz}^{T}  &  = \int_{-\infty}^{+\infty}\int_{-\infty}^{+\infty}
\partial_{x} \partial_{z} \frac{T_{12}^{TM}}{k_{1z}} e^{ik_{x}(x-x^{\prime
})+ik_{y}(y-y^{\prime})-ik_{2z}z+ik_{1z}z^{\prime}} dk_{x} dk_{y}\nonumber\\
&  = \int_{0}^{\infty}k_{s}^{2} \frac{k_{2z} T_{12}^{TM}}{k_{1z}}%
e^{-ik_{2z}z+ik_{1z}z^{\prime}} \left(  \int_{0}^{2\pi} e^{ik_{s}\rho
\cos{(\phi-\theta)}} \cos{\phi} d\phi\right)  dk_{s}\nonumber\\
&  = 2\pi i \cos{\theta} \int_{0}^{\infty}k_{s}^{2} \tilde{g}_{2,8}^{T}
J_{1}(k_{s}\rho) e^{-ik_{2z}z+ik_{1z}z^{\prime}} dk_{s}\nonumber\\
& = i(x-x^{\prime})g_{2,8}^{T},\\
{G}_{2yz}^{T}  & = \int_{-\infty}^{+\infty}\int_{-\infty}^{+\infty}
\partial_{y} \partial_{z} \frac{T_{12}^{TM}}{k_{1z}} e^{ik_{x}(x-x^{\prime
})+ik_{y}(y-y^{\prime})-ik_{2z}z+ik_{1z}z^{\prime}} dk_{x} dk_{y}\nonumber\\
& = i(y-y^{\prime})g_{2,8}^{T},
\end{align}
where
\begin{align}
g_{2,8}^{T} = 2\pi\int_{0}^{\infty} k_{s}^{2} \tilde{g}_{2,8}^{T} \frac
{J_{1}(k_{s}\rho)}{\rho} e^{-ik_{2z}z+ik_{1z}z^{\prime}} dk_{s}, \tilde
{g}_{2,8}^{T} = \frac{k_{2z} T_{12}^{TM}}{k_{1z}}.
\end{align}
The $G_{2zx}^{T}$ is derived as
\begin{align}
G_{2zx}^{T}  & = \int_{-\infty}^{+\infty}\int_{-\infty}^{+\infty} \partial_{z}
\partial_{x} \frac{T_{12}^{TM}}{k_{2z}} e^{ik_{x}(x-x^{\prime})+ik_{y}%
(y-y^{\prime})-ik_{2z}z+ik_{1z}z^{\prime}} dk_{x} dk_{y}\nonumber\\
& = \int_{0}^{\infty} k_{s}^{2} \tilde{g}_{2,9}^{T} e^{-ik_{2z}%
z+ik_{1z}z^{\prime}} \left(  \int_{0}^{2\pi} e^{ik_{s}\rho\cos{(\phi-\theta)}%
}\cos{\phi} d\phi\right) dk_{s}\nonumber\\
& = 2\pi i \cos{\theta} \int_{0}^{\infty} k_{s}^{2} \tilde{g}_{2,9}^{T}%
J_{1}(k_{s}\rho) e^{-ik_{2z}z+ik_{1z}z^{\prime}}dk_{s}\nonumber\\
&  = i(x-x^{\prime}) g_{2,9}^{T},
\end{align}
where
\begin{align}
g_{2,9}^{T} = 2\pi\int_{0}^{\infty} k_{s}^{2} \tilde{g}_{2,9}^{T} \frac
{J_{1}(k_{s}\rho)}{\rho} e^{-ik_{2z}z+ik_{1z}z^{\prime}} dk_{s}, \tilde
{g}_{2,9}^{T} = T_{12}^{TM}.
\end{align}
Similarly, $G_{2zy}^{T}$ can by found by replacing $x$ by $y$ in $G_{2zx}^{T}$ as
\begin{align}
G_{2zy}^{T}  & = \int_{-\infty}^{+\infty}\int_{-\infty}^{+\infty} \partial_{z}
\partial_{y} \frac{T_{12}^{TM}}{k_{2z}} e^{ik_{x}(x-x^{\prime})+ik_{y}%
(y-y^{\prime})-ik_{2z}z+ik_{1z}z^{\prime}} dk_{x} dk_{y} = i(y-y^{\prime})g_{2,9}^{T}.%
\end{align}

\subsection{Summary and numerical results for a two-layer structure}

The Green's function for a two-layer structure is summarized here.

\noindent$\bullet$In the first layer
\begin{align}
G_{1xx}^{R}  & =-\frac{1}{2}g_{1,5}^{R}+ \left( \frac{1}{2}\rho^{2}%
-(y-y^{\prime})^{2}\right)  g_{1,6}^{R}\label{g1xx_r}\\
G_{1yy}^{R}  & =-\frac{1}{2}g_{1,5}^{R} - \left( \frac{1}{2}\rho^{2}
-(y-y^{\prime})^{2}\right)  g_{1,6}^{R},\\
G_{1zz}^{R}  & =g_{1,7}^{R},\\
G_{1xy}^{R}  & =G_{1yx}^{R} = (x-x^{\prime})(y-y^{\prime}) g_{1,6}^{R},\\
G_{1xz}^{R}  & = -G_{1zx}^{R} = -i(x-x^{\prime})g_{1,8}^{R},\\
G_{1yz}^{R}  & = -G_{1zy}^{R} = -i(y-y^{\prime})g_{1,8}^{R}.
\end{align}
$\bullet$In the second layer
\begin{align}
G_{2xx}^{T}  & = \frac{1}{2}g_{2,5}^{T} - (\frac{1}{2}\rho^{2}-(y-y^{\prime
})^{2}) g_{2,6}^{T},\\
G_{2yy}^{T}  & = \frac{1}{2}g_{2,5}^{T} + (\frac{1}{2}\rho^{2}-(y-y^{\prime
})^{2}) g_{2,6}^{T},\\
G_{2zz}^{T}  & = g_{2,7}^{T},\\
G_{2xy}^{T}  & = -(x-x^{\prime})(y-y^{\prime})g_{2,6}^{T},\\
G_{2xz}^{T}  & =i(x-x^{\prime})g_{2,8}^{T},~~~ G_{2yz}^{T} = i(y-y^{\prime
})g_{2,8}^{T},\\
G_{2zx}^{T}  & = i(x-x^{\prime}) g_{2,9}^{T},~~~ G_{2zy}^{T} = i(y-y^{\prime})
g_{2,9}^{T}.\label{g2zx_t}%
\end{align}
\begin{figure}[t]
\centering  \includegraphics[width=4.0in]{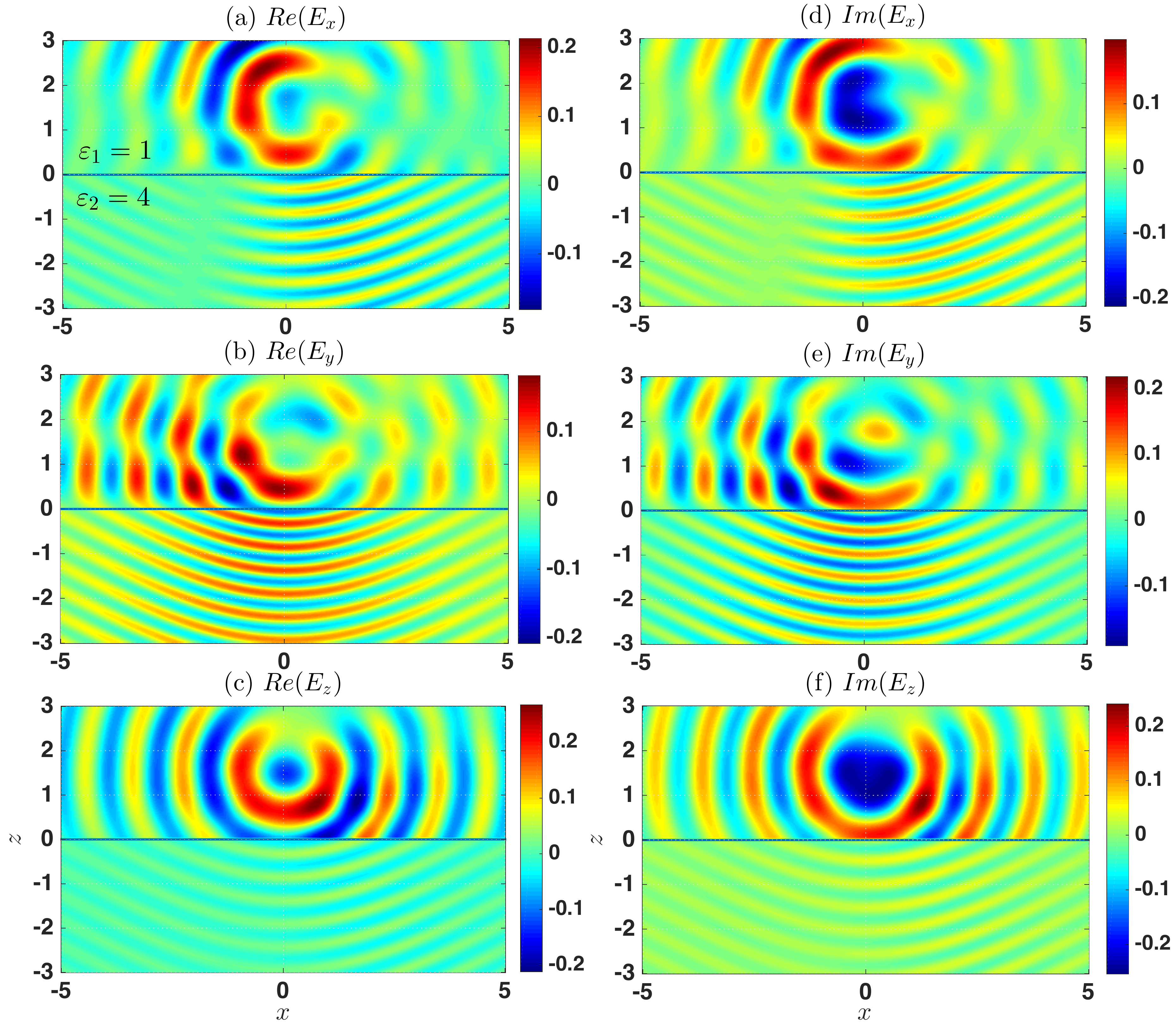}
\caption{Electric fields in a two-layer structure. A dipole source is placed at $\mathbf{r}^{\prime}= (0.1, -0.2, 1.5)$ and oriented along $\boldsymbol{\alpha}^{\prime}= (1/2, 1/2, 1/\sqrt{2})$ and Fields are computed for $-5 \leq x \leq5$ and $-3 \leq z \leq3$ for a fixed $y = 1.2$ with $\varepsilon_{1} = 1$, $\varepsilon_{2} = 4$, and $\lambda= 1$.}
\label{two_layer_field}
\end{figure}
\begin{figure}[h]
\centering
\includegraphics[width=5in]{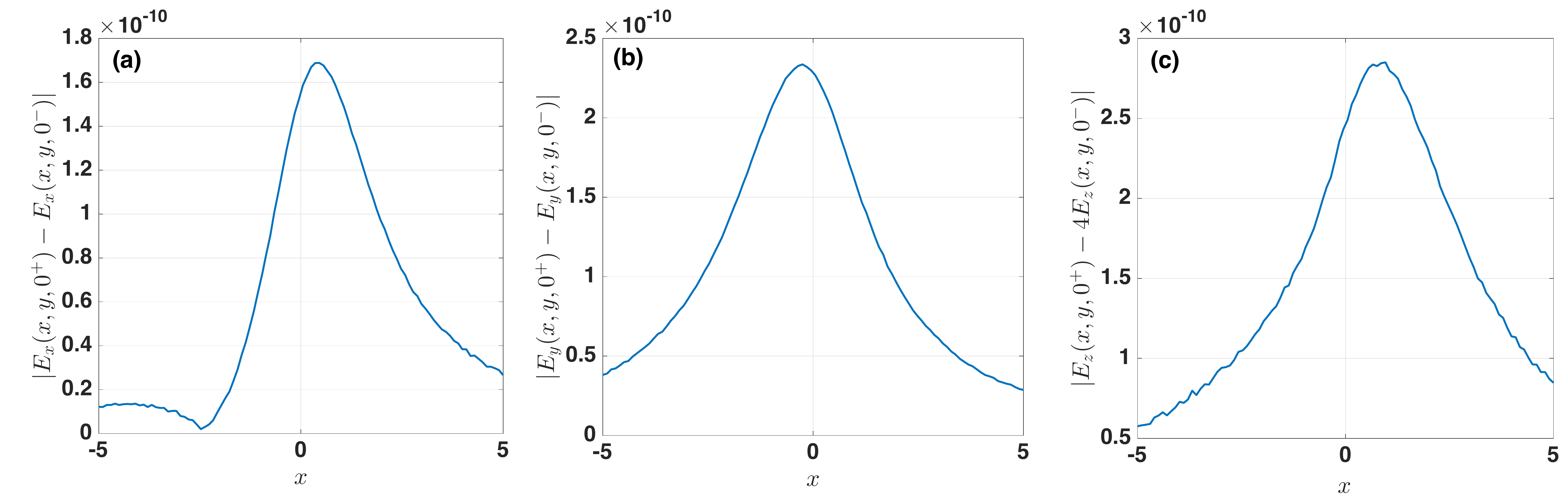}
\caption{Continuity of electric fields at interface. (a) $|E_{x}(x,y,0^{+}) -E_{x}(x,y,0^{-})|$, (b) $|E_{y}(x,y,0^{+}) -E_{y}(x,y,0^{-})|$, and (c) $|E_{z}(x,y,0^{+}) -4E_{z}(x,y,0^{-})|$. A dipole source is placed at $\mathbf{r}^{\prime}= (0.1,-0.2, 1.5)$ and oriented along $\boldsymbol{\alpha}^{\prime}= (1/2, 1/2, 1/\sqrt{2})$ and fields are computed for $-5 \leq x \leq5$ for a fixed $y =1.2$ at the layer interface $z = 0$ and with $\varepsilon_{1} = 1$, $\varepsilon_{2} = 4$, and $\lambda= 1$.}
\label{two_layer_error}
\end{figure}
Then, the dyadic Green's function for two layers is
\[
\mathbf{G}(\mathbf{r}, \mathbf{r}^{\prime}) =
\begin{cases}
\mathbf{G}^{P}- \frac{1}{8\pi^{2} \omega\varepsilon_{0} \varepsilon_{1}%
}\mathbf{G}_{1}^{R}, & z \geq0\\
~~~~~-\frac{1}{8\pi^{2} \omega\varepsilon_{0} \varepsilon_{2}}\mathbf{G}%
_{2}^{T}, & z < 0
\end{cases}
,
\]
where $\mathbf{G}^{P}$ is the free-space Green's function and each component
of $\mathbf{G}_{1}^{R}$ and $\mathbf{G}_{2}^{T}$ is given by Eqs.
(\ref{g1xx_r}) $\sim$ (\ref{g2zx_t}).

Four Sommerfeld integrals ($g_{1,5}^{R} \sim g_{1,8}^{R}$) and
five Sommerfeld integrals ($g_{1,5}^{T} \sim g_{1,9}^{T}$) are required to
compute reflected fields and transmitted parts, respectively. \emph{One less Sommerfeld integral is required than the formula presented in Ref. \cite{chew1999}}. Moreover, reflection coefficient for two layers is not assumed to reduce number of Sommerfeld integrals. As a consequence, it can be extended to multi-layered media without increasing the number of Sommerfeld integrals. Numerical integration of Sommerfeld integrals are performed with the adaptive quadrature method developed for the Helmholtz equation in Ref. \cite{gen_quad, chocai12}.

As a numerical test, a dipole source is placed at $\mathbf{r}^{\prime}= (0.1, -0.2, 1.5)$ and oriented along $\boldsymbol{\alpha}^{\prime}= (1/2, 1/2, 1/\sqrt{2})$. Then, electric field is computed for $-5 \leq x \leq5$ and $-3 \leq z \leq3$ for a fixed $y = 1.2$ with $\varepsilon_{1} = 1$, $\varepsilon_{2} = 4$, and $\lambda= 1$ in Fig. \ref{two_layer_field}. The continuity of the fields are checked by computing the electric field at the interface in Fig. \ref{two_layer_error}. First, the electric field is computed at the
layer interface $z = 0$ with the formula for the top layer, $\mathbf{E}(x,y,0^{+})$, and the formula for the bottom layer, $\mathbf{E}%
(x,y,0^{-})$. The tangential components $E_{x}$ and $E_{y}$ must be continuous and
the normal component $E_{z}$ must have a jump of $\varepsilon
_{2}/\varepsilon_{1}$ (in this example, the jump should be $4$). Figs. \ref{two_layer_error}(a), \ref{two_layer_error}(b), and \ref{two_layer_error}(c) plot $|E_{x}(x,y,0^{+}) -E_{x}(x,y,0^{-}) |$, $|E_{y}(x,y,0^{+})-E_{y}(x,y,0^{-}) |$ , and $|E_{z}(x,y,0^{+}) -4E_{z}(x,y,0^{-}) |$, respectively. About $10^{-10}$ agreement was achieved. Throughout the paper,
agreement of numerical solutions at the interface will be used as accuracy of the method. In the next section, the Green's function for a three-layer structure is presented.

\section{Green's function for a three-layer structure}

In this section, Green's function for a two-layer structure is extended to a
three-layer structure. In principle, multi-layered structure will be a
straightforward consequence of three-layer structure. We will begin the case
when a dipole source is placed in the first layer. The multiple reflection
from the second layer is accommodated with generalized Fresnel coefficients
and the continuity of the fields are ensured with by modifying
$\boldsymbol{\alpha}^{\prime}$. Note that if the source is in the second
layer, all the formulas should be reorganized to accommodate reflection from both the
first and third layer into the second layer. If the source is in the third
layer, symmetry can be used. In the next subsection, the Green's function when
the source is placed on top of a three-layer structure is derived and it is
modified to consider the case when the source is in the second layer in the
following subsection.

\subsection{Source on top of a three-layer structure}

\begin{figure}[t]
\centering  \includegraphics[width=3in]{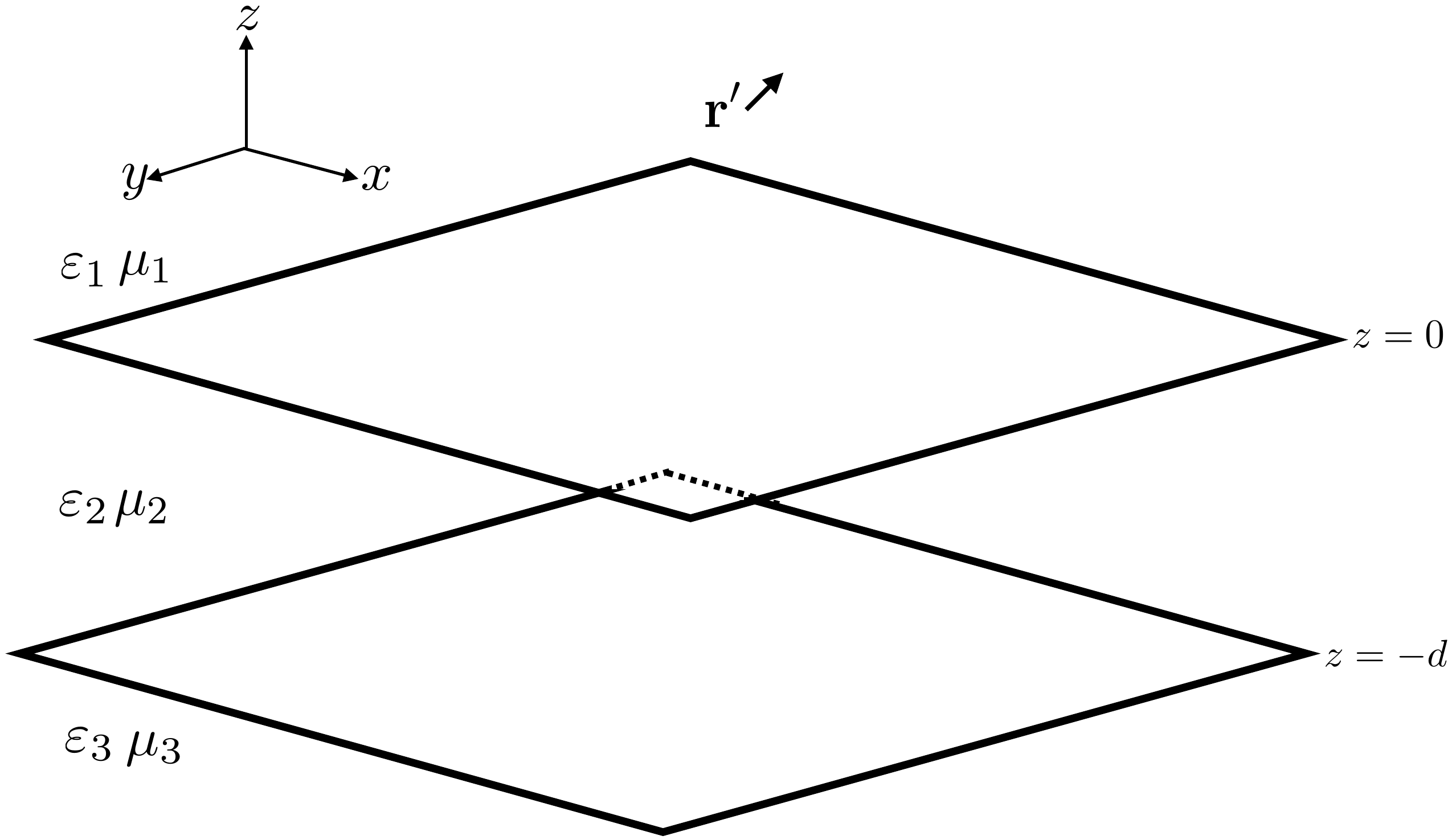}
\caption{Three-layer structure}
\label{three_layer_fig}
\end{figure}

Consider a case depicted in Fig. \ref{three_layer_fig}. A three-layer
structure is defined by two interfaces located at $z = 0$ and $z = -d$. Assume
the top most layer is the first layer with $\varepsilon_{1}$ and $\mu_{1}$, the
middle layer is the second layer with $\varepsilon_{2}$ and $\mu_{2}$, and the
bottom most layer is the third layer with $\varepsilon_{3}$ and $\mu_{3}$. Let a
dipole source is placed at $\mathbf{r}^{\prime} = (x^{\prime}, y^{\prime},
z^{\prime})$ in the first layer oriented along $\hat{\boldsymbol{\alpha}}^{\prime}= (\alpha^{\prime}_{x}, \alpha^{\prime}_{y},\alpha^{\prime}_{z})$.

\subsubsection{Green's function in the spectral domain}

The $z$-components of reflected electric and magnetic fields in the spectral domain in each layer are
\begin{align}
\tilde{E}_{1z}  & = \tilde{E}_{1z}^{P}+\tilde{E}_{1z}^{R},~\tilde{E}_{2z} =
\tilde{E}_{2z}^{R}+\tilde{E}_{2z}^{T},~ \tilde{E}_{3z} = \tilde{E}_{3z}^{T},\\
\tilde{H}_{1z}  & = \tilde{H}_{1z}^{P}+\tilde{H}_{1z}^{R},~\tilde{H}_{2z} =
\tilde{H}_{2z}^{R}+\tilde{H}_{2z}^{T},~ \tilde{H}_{3z} = \tilde{H}_{3z}^{T},
\end{align}
where the superscript $P$, $R$, and $T$ denote the primary, reflected, and
transmitted parts, respectively. The primary fields $\tilde{E}_{1z}^{P}$ and
$\tilde{H}_{1z}^{P}$ are the same as the free-space ones. The reflected part
in the first and second layers must be modified with the generalized
reflection coefficient given by
\begin{align}
\bar{R}_{12}^{TM, TE}  & = \frac{R_{12}^{TM, TE}+R^{TM, TE}_{23} e^{2ik_{2z}d}
}{1+R^{TM, TE}_{12} R^{TM, TE}_{23}e^{2ik_{2z}d}},
\end{align}
to accommodate multiple reflection and transmission from all the
layers below the first layer. Consequently, reflected fields in the first and second layer can be expressed as
\begin{align}
\tilde{E}_{1z}^{R}  & = \left(  \hat{z} \cdot\hat{\boldsymbol{\alpha}}%
_{1R}^{\prime\prime}+\frac{1}{k_{1}^{2}} \partial_{z} \nabla\cdot
\hat{\boldsymbol{\alpha}}_{1R}^{\prime\prime}\right)  \tilde{g}_{1, TM}^{R},~
\tilde{H}_{1z}^{R} = -\frac{1}{i\omega\mu_{0}\mu_{1}}\hat{z} \cdot\nabla_{s}
\times\hat{\boldsymbol{\alpha}}_{1R}^{\prime\prime}\tilde{g}_{1,TE}%
^{R},\label{ejz_r}\\
\tilde{E}_{2z}^{R}  & = \left(  \hat{z} \cdot\hat{ \boldsymbol{\alpha}}%
_{2R}^{\prime\prime}+\frac{1}{k_{2}^{2}} \partial_{z} \nabla\cdot
\hat{\boldsymbol{\alpha}}_{2R}^{\prime\prime}\right)  \tilde{g}_{2, TM}^{R},~
\tilde{H}_{2z}^{R} = -\frac{1}{i\omega\mu_{0}\mu_{2}}\frac{k_{2z}}{k_{1z}}%
\hat{z} \cdot\nabla_{s} \times\hat{\boldsymbol{\alpha}}_{2R}^{\prime\prime
}\tilde{g}_{2,TE}^{R},\label{ejz_r}%
\end{align}
where
\begin{align}
& \tilde{g}_{1, TM,TE}^{R} = -R_{1}^{TM, TE} \frac{\omega\mu_{0}\mu_{1}}%
{8\pi^{2}} \frac{e^{ik_{x}(x-x^{\prime})+ik_{y}(y-y^{\prime})+ik_{1z}%
(z+z^{\prime})}}{k_{1z}},\\
& \tilde{g}_{2, TM,TE}^{R} = -R_{2}^{TM, TE}\frac{\omega\mu_{0}\mu_{2}}%
{8\pi^{2}} \frac{e^{ik_{x}(x-x^{\prime})+ik_{y}(y-y^{\prime})+ik_{2z}%
z+ik_{1z}z^{\prime}+2ik_{2z} d}}{k_{1z}},\\
& \hat{\boldsymbol{\alpha}}_{1R}^{\prime\prime}= (-\alpha_{x}^{\prime},
-\alpha_{y}^{\prime}, \alpha_{z}^{\prime}), \hat{\boldsymbol{\alpha}}%
_{2R}^{\prime\prime}= (-\frac{k_{1z}}{k_{2z}}\alpha_{x}^{\prime},
-\frac{k_{1z}}{k_{2z}}\alpha_{y}^{\prime}, \alpha_{z}^{\prime}),\\
& R_{1}^{TM, TE} = \bar{R}_{12}^{TM, TE},~ R_{2}^{TM, TE} = A_{2}^{TM,
TE}R_{23}^{TM, TE}, A_{2}^{TM, TE} = \frac{T_{12}^{TM, TE}}{1-R_{21}^{TM,
TE}R_{23}^{TM, TE}e^{2ik_{2z}d}}.\label{new_ref_coef}%
\end{align}
The transmitted parts in the second and third layer must be modified to
\begin{align}
\tilde{E}_{jz}^{T}  & = \left(  \hat{z} \cdot\hat{\alpha}_{jT}^{\prime
\prime\prime}+ \frac{1}{k^{2}_{j}} \partial_{z} \nabla\cdot\hat{\alpha}%
_{jT}^{\prime\prime\prime}\right)  g_{j, TM}^{T},~ \tilde{H}_{jz}^{T} =
\frac{1}{i\omega\mu_{0} \mu_{j} }\frac{\mu_{1}}{\mu_{j}} \frac{k_{jz}}{k_{1z}}
\hat{z} \cdot\nabla_{s} \times\hat{\alpha}^{\prime\prime\prime}_{jT} g_{j,
TE}^{T}, j=2,3,
\end{align}
where
\begin{align}
& \tilde{g}_{2, TM,TE}^{T} = -A_{2}^{TM, TE} \frac{\omega\mu_{0}\mu_{2}}%
{8\pi^{2}} \frac{e^{ik_{x}(x-x^{\prime})+ik_{y}(y-y^{\prime})-ik_{2z}%
z+ik_{1z}z^{\prime}}}{k_{1z}},\\
& \tilde{g}_{3, TM,TE}^{T} = -A_{3}^{TM, TE}\frac{\omega\mu_{0}\mu_{3}}%
{8\pi^{2}} \frac{e^{ik_{x}(x-x^{\prime})+ik_{y}(y-y^{\prime})-ik_{3z}%
z+ik_{1z}z^{\prime}}}{k_{1z}},\\
& \alpha_{jT}^{\prime\prime\prime}= \left( \frac{k_{1z}}{k_{jz}}\alpha
_{x}^{\prime}, \frac{k_{1z}}{k_{jz}}\alpha_{y}^{\prime}, \alpha_{z}^{\prime
}\right) , j=2,3,\\
& A_{3}^{TM, TE} = A_{2}^{TM, TE}(1+R_{23}^{TM, TE}) e^{ik_{2z}d-ik_{3z}d}.\label{new_trans_coef}%
\end{align}
All the reflection and transmission coefficients are changed to enforce
multiple reflections in Eqs. (\ref{new_ref_coef}) and (\ref{new_trans_coef}).
At the same time, in each layer $\boldsymbol{\alpha}^{\prime}$ is modified to
correctly ensure the continuity of the fields at the interfaces. The transverse
components can be derived using Eq. (\ref{E_1s}) using the new $z$-components
listed above in each layer. In the following, all components are presented in each layer.\newline

\noindent$\bullet$ In the first layer, the transverse components of reflected
parts is found by
\begin{align}
\tilde{\mathbf{E}}_{1s}^{R}  & = \frac{1}{k_{s}^{2}}\left[ \nabla_{s}
\partial_{z} \tilde{E}_{1z}^{R} - i\omega\mu_{0} \mu_{1} \hat{z} \times
\nabla_{s} \tilde{H}_{1z}^{R} \right] \nonumber\\
&  = \frac{1}{k_{s}^{2}}\nabla_{s} \partial_{z} \left(  \hat{z} \cdot
\hat{\alpha}_{1R}^{\prime\prime}+\frac{1}{k_{1}^{2}} \partial_{z} \nabla
\cdot\hat{\alpha}_{1R}^{\prime\prime}\right)  \tilde{g}_{1, TM}^{R}+\frac
{1}{k_{s}^{2}}\hat{z} \times\nabla_{s} \left( \hat{z} \cdot\nabla_{s}
\times\hat{\alpha}_{1R}^{\prime\prime}\tilde{g}_{1,TE}^{R}\right) .
\end{align}
Each component is exactly the same as the two-layer structure except the
definition of the reflection coefficients. Thus the reflected parts of the
Green's function in the first layer can be simply rewritten by replacing
$R_{12}^{TM}$ and $R_{12}^{TE}$ by the generalized reflection coefficient $R_{1}^{TM} = \tilde{R}_{12}^{TM}$ and $R_{1}^{TE} = \tilde{R}_{12}^{TE}$, respectively. Therefore, the electric field in the spectral domain in the first layer is
\begin{align}
\left[
\begin{array}
[c]{c}%
\tilde{E}_{1x}\\
\tilde{E}_{1y}\\
\tilde{E}_{1z}%
\end{array}
\right]   & = \left[
\begin{array}
[c]{c}%
\tilde{E}_{1x}^{P}+\tilde{E}_{1x}^{R}\\
\tilde{E}_{1y}^{P}+\tilde{E}_{1y}^{R}\\
\tilde{E}_{1z}^{P}+\tilde{E}_{1z}^{R}%
\end{array}
\right]  =\left( \tilde{\mathbf{G}}^{P}-\frac{1}{8\pi^{2}\omega\varepsilon
_{0}\varepsilon_{1}}\tilde{\mathbf{G}}_{1}^{R}\right)  \boldsymbol{\alpha}^{\prime
}\nonumber\\
& = \left( \left[
\begin{array}
[c]{ccc}%
\tilde{G}^{P}_{xx} & \tilde{G}^{P}_{xy} & \tilde{G}^{P}_{xz}\\
\tilde{G}^{P}_{yx} & \tilde{G}^{P}_{yy} & \tilde{G}^{P}_{yz}\\
\tilde{G}^{P}_{zx} & \tilde{G}^{P}_{zy} & \tilde{G}^{P}_{zz}%
\end{array}
\right]  - \frac{1}{8\pi^{2}\omega\varepsilon_{0}\varepsilon_{1}}\left[
\begin{array}
[c]{ccc}%
\tilde{G}^{R}_{1xx} & \tilde{G}^{R}_{1xy} & \tilde{G}^{R}_{1xz}\\
\tilde{G}^{R}_{1yx} & \tilde{G}^{R}_{1yy} & \tilde{G}^{R}_{1yz}\\
\tilde{G}^{R}_{1zx} & \tilde{G}^{R}_{1zy} & \tilde{G}^{R}_{1zz}%
\end{array}
\right]  \right) \left[
\begin{array}
[c]{c}%
\alpha_{x}^{\prime}\\
\alpha_{y}^{\prime}\\
\alpha_{z}^{\prime}%
\end{array}
\right] ,
\end{align}
where $\tilde{\mathbf{G}}^{P}$ is the same as Eq. (\ref{spectral_free}) and
$\tilde{\mathbf{G}}^{R}_{1}$ is defined by
\begin{align}
\tilde{G}^{R}_{1xx} & = \left(  \partial_{x}^{2} \frac{k_{1z}}{k_{s}^{2}}%
R_{1}^{TM}-\partial_{y}^{2} \frac{k_{1}^{2}}{k_{s}^{2} k_{1z}}R_{1}^{TE}
\right) e^{ik_{x}(x-x^{\prime})+ik_{y}(y-y^{\prime})+ik_{1z}(z+z^{\prime}%
)},\label{1xx_2}\\
\tilde{G}^{R}_{1yy} & = \left(  \partial_{y}^{2} \frac{k_{1z}}{k_{s}^{2}}%
R_{1}^{TM}-\partial_{x}^{2} \frac{k_{1}^{2}}{k_{s}^{2} k_{1z}}R_{1}^{TE}
\right) e^{ik_{x}(x-x^{\prime})+ik_{y}(y-y^{\prime})+ik_{1z}(z+z^{\prime}%
)},\label{1yy_2}\\
\tilde{G}_{1zz}^{R}  &  = \left(  \frac{k_{s}^{2}}{k_{1z}} R_{1}^{TM}\right)
e^{ik_{x}(x-x^{\prime})+ik_{y}(y-y^{\prime})+ik_{1z}(z+z^{\prime})}
,\label{1zz_2}\\
\tilde{G}_{1xy}^{R} & =\tilde{G}_{1yx}^{R} = \partial_{x}\partial_{y} \left(
\frac{k_{1z}}{k_{s}^{2} }R_{1}^{TM}+\frac{k_{1}^{2}}{k_{s}^{2}k_{1z}}%
R_{1}^{TE}\right)  e^{ik_{x}(x-x^{\prime})+ik_{y}(y-y^{\prime})+ik_{1z}%
(z+z^{\prime})},\label{1xy_2}\\
\tilde{G}_{1xz}^{R}  & = -\tilde{G}_{1zx}^{R}= \partial_{x} \partial_{z}
\left(  \frac{R_{1}^{TM}}{k_{1z}} \right)  e^{ik_{x}(x-x^{\prime}%
)+ik_{y}(y-y^{\prime})+ik_{1z}(z+z^{\prime})} ,\label{1xz_2}\\
\tilde{G}_{1yz}^{R}  & = -\tilde{G}_{1zy}^{R} = \partial_{y} \partial_{z}
\left(  \frac{R_{1}^{TM}}{k_{1z}} \right)  e^{ik_{x}(x-x^{\prime}%
)+ik_{y}(y-y^{\prime})+ik_{1z}(z+z^{\prime})} .\label{1yz_2}%
\end{align}
\noindent$\bullet$ In the second layer, there are both reflected and
transmitted parts ($\mathbf{\tilde{E}}_{2} = \mathbf{\tilde{E}}^{R}_{2}+\mathbf{\tilde{E}}^{T}_{2}$),
\begin{align}
\left[
\begin{array}
[c]{c}%
\tilde{E}_{2x}\\
\tilde{E}_{2y}\\
\tilde{E}_{2z}%
\end{array}
\right]   & = \left[
\begin{array}
[c]{c}%
\tilde{E}_{2x}^{R}+\tilde{E}_{2x}^{T}\\
\tilde{E}_{2y}^{R}+\tilde{E}_{2y}^{T}\\
\tilde{E}_{2z}^{R}+\tilde{E}_{2z}^{T}%
\end{array}
\right]  =-\frac{1}{8\pi^{2}\omega\varepsilon_{0}\varepsilon_{2}} \left(
\tilde{\mathbf{G}}_{2}^{R}+\tilde{\mathbf{G}}_{2}^{T}\right)  \boldsymbol{\alpha}^{\prime
}\nonumber\\
& = -\frac{1}{8\pi^{2}\omega\varepsilon_{0}\varepsilon_{2}} \left( \left[
\begin{array}
[c]{ccc}%
\tilde{G}^{R}_{2xx} & \tilde{G}^{R}_{2xy} & \tilde{G}^{R}_{2xz}\\
\tilde{G}^{R}_{2yx} & \tilde{G}^{R}_{2yy} & \tilde{G}^{R}_{2yz}\\
\tilde{G}^{R}_{2zx} & \tilde{G}^{R}_{2zy} & \tilde{G}^{R}_{2zz}%
\end{array}
\right]  + \left[
\begin{array}
[c]{ccc}%
\tilde{G}^{T}_{2xx} & \tilde{G}^{T}_{2xy} & \tilde{G}^{T}_{2xz}\\
\tilde{G}^{T}_{2yx} & \tilde{G}^{T}_{2yy} & \tilde{G}^{T}_{2yz}\\
\tilde{G}^{T}_{2zx} & \tilde{G}^{T}_{2zy} & \tilde{G}^{T}_{2zz}%
\end{array}
\right]  \right) \left[
\begin{array}
[c]{c}%
\alpha_{x}^{\prime}\\
\alpha_{y}^{\prime}\\
\alpha_{z}^{\prime}%
\end{array}
\right] .
\end{align}
The transmitted part assumes the same form as for the two-layer case. Thus,
the transmitted part of the Green's function can be simply found by replacing
the transmission coefficient $T_{12}^{TM, TE}$ by the $A_{2}^{TM, TE}$ as
\begin{align}
\tilde{G}_{2xx}^{T}  & = \left( -\partial_{x}^{2} \frac{k_{2z}}{k_{s}^{2}}
A_{2}^{TM}-\partial_{y}^{2} \frac{k_{2}^{2} \mu_{1}}{k_{s}^{2} \mu_{2} k_{1z}%
}A_{2}^{TE}\right)  e^{ik_{x}(x-x^{\prime})+ik_{y}(y-y^{\prime})-ik_{2z}z
+ik_{1z}z^{\prime}},\\
\tilde{G}_{2yy}^{T}  & = \left( -\partial_{y}^{2} \frac{k_{2z}}{k_{s}^{2}}
A_{2}^{TM} - \partial_{x}^{2}\frac{k_{2}^{2} \mu_{1}}{k_{s}^{2} \mu_{2}
k_{1z}}A_{2}^{TE} \right) e^{ik_{x}(x-x^{\prime})+ik_{y}(y-y^{\prime}%
)-ik_{2z}z +ik_{1z}z^{\prime}},\\
\tilde{G}_{2zz}^{T}  & = \left(  \frac{k_{s}^{2}}{k_{1z}}A_{2}^{TM} \right)
e^{ik_{x}(x-x^{\prime})+ik_{y}(y-y^{\prime})-ik_{2z}z +ik_{1z}z^{\prime}},\\
\tilde{G}_{2xy}^{T}  & = \tilde{G}_{2yx}^{T} = \left( -\partial_{x}
\partial_{y} \frac{k_{2z}}{k_{s}^{2}}A_{2}^{TM}+\partial_{x} \partial_{y}
\frac{k_{2}^{2} \mu_{1}}{k_{s}^{2} \mu_{2} k_{1z}}A_{2}^{TE} \right)
e^{ik_{x}(x-x^{\prime})+ik_{y}(y-y^{\prime})-ik_{2z}z +ik_{1z}z^{\prime}},\\
\tilde{G}_{2xz}^{T}  & = \partial_{x} \partial_{z} \frac{A_{2}^{TM}}{k_{1z}%
}e^{ik_{x}(x-x^{\prime})+ik_{y}(y-y^{\prime})-ik_{2z}z +ik_{1z}z^{\prime}%
},\tilde{G}_{2yz}^{T} = \partial_{y} \partial_{z} \frac{A_{2}^{TM}}{k_{1z}%
}e^{ik_{x}(x-x^{\prime})+ik_{y}(y-y^{\prime})-ik_{2z}z +ik_{1z}z^{\prime}},\\
\tilde{G}_{2zx}^{T}  & = \partial_{z} \partial_{x} \frac{A_{2}^{TM}}{k_{2z}%
}e^{ik_{x}(x-x^{\prime})+ik_{y}(y-y^{\prime})-ik_{2z}z +ik_{1z}z^{\prime}%
},\tilde{G}_{2zy}^{T} = \partial_{z} \partial_{y} \frac{A_{2}^{TM}}{k_{2z}%
}e^{ik_{x}(x-x^{\prime})+ik_{y}(y-y^{\prime})-ik_{2z}z +ik_{1z}z^{\prime}}.
\end{align}
However, there are some changes in the reflected parts in the second layer
since the definition of $\hat{\boldsymbol{\alpha}}^{\prime\prime}$ of the
reflected parts in the first layer is changed to $\hat{\boldsymbol{\alpha}}%
_{2R}^{\prime\prime}$. Fortunately, the reflected part takes a similar form as the
transmitted part because of $-k_{1z}/k_{2z}$ in $\hat{\boldsymbol{\alpha}}_{2R}^{\prime\prime}$. By carefully re-deriving the transverse component of
reflected parts in the second layer, the spectral Green's function can be found as
\begin{align}
\tilde{G}_{2xx}^{R}  & = \left(  \partial_{x}^{2} \frac{k_{2z}}{k_{s}^{2}}
R_{2}^{TM} -\partial_{y}^{2} \frac{k_{2}^{2}}{k_{s}^{2} k_{1z}} R_{2}^{TE}
\right)  e^{ik_{x}(x-x^{\prime})+ik_{y}(y-y^{\prime})+ik_{2z}z+ik_{1}%
z^{\prime}+2ik_{2z}d},\\
\tilde{G}_{2yy}^{R}  & = \left(  \partial_{y}^{2} \frac{k_{2z}}{k_{s}^{2}}
R_{2}^{TM} -\partial_{x}^{2} \frac{k_{2}^{2}}{k_{s}^{2} k_{1z}} R_{2}^{TE}
\right)  e^{ik_{x}(x-x^{\prime})+ik_{y}(y-y^{\prime})+ik_{2z}z+ik_{1}%
z^{\prime}+2ik_{2z}d},\\
\tilde{G}_{2zz}^{R}  & = \left( \frac{k_{s}^{2}}{k_{1z}}R_{2}^{TM} \right)
e^{ik_{x}(x-x^{\prime})+ik_{y}(y-y^{\prime})+ik_{2z}z+ik_{1}z^{\prime
}+2ik_{2z}d},\\
\tilde{G}_{2xy}^{R}  & = \tilde{G}_{2yx}^{R} = \left( \partial_{x}
\partial_{y} \frac{k_{2z}}{k_{s}^{2}}R_{2}^{TM}+\partial_{x}\partial_{y}
\frac{k_{2}^{2}}{k_{s}^{2} k_{1z}}R_{2}^{TE} \right) e^{ik_{x}(x-x^{\prime
})+ik_{y}(y-y^{\prime})+ik_{2z}z+ik_{1}z^{\prime}+2ik_{2z}d},\\
\tilde{G}_{2xz}^{R}  & = \partial_{x} \partial_{z} \frac{R_{2}^{TM}}{k_{1z}%
}e^{ik_{x}(x-x^{\prime})+ik_{y}(y-y^{\prime})+ik_{2z}z+ik_{1}z^{\prime
}+2ik_{2z}d},\tilde{G}_{2yz}^{R} = \partial_{y}\partial_{z} \frac{R_{2}^{TM}%
}{k_{1z}}e^{ik_{x}(x-x^{\prime})+ik_{y}(y-y^{\prime})+ik_{2z}z+ik_{1}%
z^{\prime}+2ik_{2z}d},\\
\tilde{G}_{2zx}^{R}  & = -\partial_{x}\partial_{z} \frac{R_{2}^{TM}}{k_{2z}%
}e^{ik_{x}(x-x^{\prime})+ik_{y}(y-y^{\prime})+ik_{2z}z+ik_{1}z^{\prime
}+2ik_{2z}d}, \tilde{G}_{2zy}^{R} = -\partial_{y}\partial_{z} \frac{R_{2}%
^{TM}}{k_{2z}}e^{ik_{x}(x-x^{\prime})+ik_{y}(y-y^{\prime})+ik_{2z}%
z+ik_{1}z^{\prime}+2ik_{2z}d}.
\end{align}

\noindent$\bullet$ In the third layer, the transverse component of the transmitted part takes the same form as the transmitted part of the second layer. They can be simply found by changing the index in the transmitted part of the second layer. By combining all the components, the spectral Green's function in the
third layer can be expressed by
\begin{align}
\left[
\begin{array}
[c]{c}%
\tilde{E}_{3x}\\
\tilde{E}_{3y}\\
\tilde{E}_{3z}%
\end{array}
\right]   & = \left[
\begin{array}
[c]{c}%
\tilde{E}_{3x}^{T}\\
\tilde{E}_{3y}^{T}\\
\tilde{E}_{3z}^{T}%
\end{array}
\right]  =-\frac{1}{8\pi^{2}\omega\varepsilon_{0}\varepsilon_{3}}%
\tilde{\mathbf{G}}_{2}^{T} \boldsymbol{\alpha}^{\prime}= -\frac{1}{8\pi^{2}\omega
\varepsilon_{0}\varepsilon_{3}} \left[
\begin{array}
[c]{ccc}%
\tilde{G}^{T}_{3xx} & \tilde{G}^{T}_{3xy} & \tilde{G}^{T}_{3xz}\\
\tilde{G}^{T}_{3yx} & \tilde{G}^{T}_{3yy} & \tilde{G}^{T}_{3yz}\\
\tilde{G}^{T}_{3zx} & \tilde{G}^{T}_{3zy} & \tilde{G}^{T}_{3zz}%
\end{array}
\right]  \left[
\begin{array}
[c]{c}%
\alpha_{x}^{\prime}\\
\alpha_{y}^{\prime}\\
\alpha_{z}^{\prime}%
\end{array}
\right] ,
\end{align}
where
\begin{align}
\tilde{G}_{3xx}^{T}  &  = \left( -\partial_{x}^{2} \frac{k_{3z}}{k_{s}^{2}%
}A_{3}^{TM}-\partial_{y}^{2} \frac{k_{3}^{2} \mu_{1}}{k_{s}^{2} \mu_{3}
k_{1z}}A_{3}^{TE} \right)  e^{ik_{x}(x-x^{\prime})+ik_{y}(y-y^{\prime
})-ik_{3z}z+ik_{1z}z^{\prime}},\\
\tilde{G}_{3yy}^{T}  &  = \left( -\partial_{y}^{2} \frac{k_{3z}}{k_{s}^{2}%
}A_{3}^{TM} - \partial_{x}^{2} \frac{k_{3}^{2} \mu_{1}}{k_{s}^{2} \mu_{3}
k_{1z}}A_{3}^{TE} \right) e^{ik_{x}(x-x^{\prime})+ik_{y}(y-y^{\prime}%
)-ik_{3z}z+ik_{1z}z^{\prime}},\\
\tilde{G}_{3zz}^{T}  &  = \frac{k_{s}^{2}}{k_{1z}}A_{3}^{TM}e^{ik_{x}%
(x-x^{\prime})+ik_{y}(y-y^{\prime})-ik_{3z}z+ik_{1z}z^{\prime}},\\
\tilde{G}_{3xy}^{T}  &  =\tilde{G}_{3yx}^{T} = \left( -\partial_{x}
\partial_{y} \frac{k_{3z}}{k_{s}^{2}} A_{3}^{TM}+\partial_{x} \partial_{y}
\frac{k_{3}^{2} \mu_{1}}{k_{s}^{2} \mu_{3} k_{1z}}A_{3}^{TE} \right)
e^{ik_{x}(x-x^{\prime})+ik_{y}(y-y^{\prime})-ik_{3z}z+ik_{1z}z^{\prime}},\\
\tilde{G}_{3xz}^{T}  &  = \partial_{x} \partial_{z} \frac{A_{3}^{TM}}{k_{1z}%
}e^{ik_{x}(x-x^{\prime})+ik_{y}(y-y^{\prime})-ik_{3z}z+ik_{1z}z^{\prime}},
\tilde{G}_{3yz}^{T} = \partial_{y} \partial_{z} \frac{A_{3}^{TM}}{k_{1z}%
}e^{ik_{x}(x-x^{\prime})+ik_{y}(y-y^{\prime})-ik_{3z}z+ik_{1z}z^{\prime}},\\
\tilde{G}_{3zx}^{T}  &  = \partial_{z} \partial_{x} \frac{A_{3}^{TM}}{k_{3z}%
}e^{ik_{x}(x-x^{\prime})+ik_{y}(y-y^{\prime})-ik_{3z}z+ik_{1z}z^{\prime}},
\tilde{G}_{3zy}^{T} = \partial_{z} \partial_{y} \frac{A_{3}^{TM}}{k_{3z}%
}e^{ik_{x}(x-x^{\prime})+ik_{y}(y-y^{\prime})-ik_{3z}z+ik_{1z}z^{\prime}}.
\end{align}

\subsubsection{Green's function in the spatial domain}

The inverse Fourier transform is taken to recover the Green's function in the spatial domain as before. In the spectral domain, the reflected part in the first
layer and transmitted part in the second and third layer have the exactly same
form as the two-layer Green's function. Thus, the spatial domain Green's
function can be simply found by replacing the reflection and transmission
coefficient and index without actual derivation. The reflected part in the
second layer has almost identical form as the transmitted part in the second
layer due to similar definition of $\boldsymbol{\alpha}^{\prime\prime}_{2R} =
(-\frac{k_{1z}}{k_{2z}}\alpha_{x}^{\prime}, -\frac{k_{1z}}{k_{2z}}\alpha
_{y}^{\prime}, \alpha_{z}^{\prime})$ and $\boldsymbol{\alpha}^{\prime\prime
}_{2T} = (\frac{k_{1z}}{k_{2z}}\alpha_{x}^{\prime}, \frac{k_{1z}}{k_{2z}%
}\alpha_{y}^{\prime}, \alpha_{z}^{\prime})$. Therefore, by carefully changing
the sign of transmitted part Green's function formula, one can find the reflected
part in the second layer.\newline

\noindent$\bullet$ In the first layer, the reflected part Green's function is
given by
\begin{align}
G_{1xx}^{R}  & =-\frac{1}{2}g_{1,5}^{R}+ \left( \frac{1}{2}\rho^{2}%
-(y-y^{\prime})^{2}\right)  g_{1,6}^{R},\\
G_{1yy}^{R}  & =-\frac{1}{2}g_{1,5}^{R} - \left( \frac{1}{2}\rho^{2}
-(y-y^{\prime})^{2}\right)  g_{1,6}^{R},\\
G_{1zz}^{R}  & =g_{1,7}^{R},\\
G_{1xy}^{R}  & =G_{1yx}^{R} = (x-x^{\prime})(y-y^{\prime}) g_{1,6}^{R},\\
G_{1xz}^{R}  & = -G_{1zx}^{R} = -i(x-x^{\prime})g_{1,8}^{R},\\
G_{1yz}^{R}  & = -G_{1zy}^{R} = -i(y-y^{\prime})g_{1,8}^{R},
\end{align}
where
\begin{align}
\tilde{g}_{1,5}^{R}  &  =k_{1z}R_{1}^{TM}-\frac{k_{1}^{2}}{k_{1z}}R_{1}^{TE}
,\tilde{g}_{1,6}^{R} =\frac{k_{1z}}{k_{s}^{2}}R_{1}^{TM}+\frac{k_{1}^{2}%
}{k_{s}^{2}k_{1z}}R_{1}^{TE}, \tilde{g}_{1,7}^{R} = \frac{k_{s}^{2}}{k_{1z}}
R_{1}^{TM}, \tilde{g}_{1,8}^{R} = R_{1}^{TM},\nonumber\\
g_{1,5}^{R}  & = 2\pi\int_{0}^{\infty} k_{s} \tilde{g}_{1,5}^{R} J_{0}%
(k_{s}\rho) e^{ik_{1z}(z+z^{\prime})}dk_{s}, g_{1,6}^{R} = 2\pi\int
_{0}^{\infty} k_{s}^{3} \tilde{g}_{1,6}^{R} \frac{J_{2}(k_{s}\rho)}{\rho^{2}%
}e^{ik_{1}z(z+z^{\prime})}dk_{s},\nonumber\\
g_{1,7}^{R}  & = 2\pi\int_{0}^{\infty} k_{s} \tilde{g}_{1,7}^{R} J_{0}%
(k_{s}\rho) e^{ik_{1z}(z+z^{\prime})} dk_{s}, g_{1,8}^{R} = 2\pi\int
_{0}^{\infty}k^{2}_{s} \tilde{g}_{1,8}^{R} \frac{J_{1}(k_{s}\rho)}{\rho}
e^{ik_{1z}(z+z^{\prime})} dk_{s}.
\end{align}

\noindent$\bullet$ In the second layer, the transmitted part has the same form
as the two layers case. Therefore, the Green's function can be found simply
replacing $T_{12}^{TM}$ and $T_{12}^{TE}$ by $A_{2}^{TM}$ and $A_{2}^{TE}$ as
\begin{align}
G_{2xx}^{T}  & = \frac{1}{2}g_{2,5}^{T} - (\frac{1}{2}\rho^{2}-(y-y^{\prime
})^{2}) g_{2,6}^{T},\\
G_{2yy}^{T}  & = \frac{1}{2}g_{2,5}^{T} + (\frac{1}{2}\rho^{2}-(y-y^{\prime
})^{2}) g_{2,6}^{T},\\
G_{2zz}^{T}  & = g_{2,7}^{T},\\
G_{2xy}^{T}  & = G_{2yx}^{T} = -(x-x^{\prime})(y-y^{\prime})g_{2,6}^{T},\\
G_{2xz}^{T}  & =i(x-x^{\prime})g_{2,8}^{T},~~~ G_{2yz}^{T} = i(y-y^{\prime
})g_{2,8}^{T},\\
G_{2zx}^{T}  & = i(x-x^{\prime}) g_{2,9}^{T},~~~ G_{2zy}^{T} = i(y-y^{\prime})
g_{2,9}^{T},
\end{align}
where
\begin{align}
& \tilde{g}_{2,5}^{T} = k_{2z} A_{2}^{TM}+\frac{k_{2}^{2}}{k_{1z}}\frac
{\mu_{1}}{\mu_{2}} A_{2}^{TE}, \tilde{g}_{2,6}^{T}= \frac{k_{2z} }{k_{s}^{2}%
}A_{2}^{TM} - \frac{k_{2}^{2}}{k_{1z}k_{s}^{2}}\frac{\mu_{1}}{\mu_{2}}
A_{2}^{TE},\nonumber\\
& \tilde{g}_{2,7}^{T} = \frac{k_{s}^{2}}{k_{1z}} A_{2}^{TM},\tilde{g}%
_{2,8}^{T} = \frac{k_{2z} A_{2}^{TM}}{k_{1z}} , \tilde{g}_{2,9}^{T} =
A_{2}^{TM},\nonumber\\
& g_{2,5}^{T} = 2\pi\int_{0}^{\infty} k_{s} \tilde{g}_{2,5}^{T}J_{0}(k_{s}%
\rho) e^{-ik_{2z}z+ik_{1z}z^{\prime}}dk_{s}, g_{2,6}^{T} = 2\pi\int
_{0}^{\infty} k_{s}^{3} \tilde{g}_{2,6} \frac{J_{2}(k_{s}\rho)}{\rho^{2}}^{T}
e^{-ik_{2z}z+ik_{1z}z^{\prime}}dk_{s},\nonumber\\
& g_{2,7}^{T} = 2\pi\int_{0}^{\infty} k_{s} \tilde{g}_{2,7}^{T} J_{0}%
(k_{s}\rho) e^{-ik_{2z}z+ik_{1z}z^{\prime}} dk_{s}, g_{2,8}^{T} = 2\pi\int
_{0}^{\infty}k^{2}_{s} \tilde{g}_{2,8}^{T} \frac{J_{1}(k_{s}\rho)}{\rho}
e^{-ik_{2z}z+ik_{1z}z^{\prime}} dk_{s},\nonumber\\
& g_{2,9}^{T} = 2\pi\int_{0}^{\infty}k^{2}_{s} \tilde{g}_{2,8}^{T} \frac
{J_{1}(k_{s}\rho)}{\rho} e^{-ik_{2z}z+ik_{1z}z^{\prime}} dk_{s}.
\end{align}
The reflected part in the second layer is derived by observing the similarity between the transmitted part and reflected parts in the second layer, one can change the sign of
transmitted part to obtain the Green's function or one can take actual double integral and derive the same formulas.
\begin{align}
G_{2xx}^{R}  & =-\frac{1}{2}g_{2,5}^{R}+ \left( \frac{1}{2}\rho^{2}%
-(y-y^{\prime})^{2}\right)  g_{2,6}^{R},\\
G_{2yy}^{R}  & =-\frac{1}{2}g_{2,5}^{R} - \left( \frac{1}{2}\rho^{2}
-(y-y^{\prime})^{2}\right)  g_{2,6}^{R},\\
G_{2zz}^{R}  & =g_{2,7}^{R},\\
G_{2xy}^{R}  & =G_{2yx}^{R} = (x-x^{\prime})(y-y^{\prime}) g_{2,6}^{R},\\
G_{2zx}^{R}  & = i(x-x^{\prime})g_{2,8}^{R}, G_{2zy}^{R} = i(y-y^{\prime
})g_{2,8}^{R},\\
G_{2xz}^{R}  & = -i(x-x^{\prime})g_{2,9}^{R}, G_{2yz}^{R} = -i(y-y^{\prime
})g_{2,9}^{R},
\end{align}
where
\begin{align}
& \tilde{g}_{2,5}^{R} =k_{2z}R_{2}^{TM}-\frac{k_{2}^{2}}{k_{1z}}R_{2}^{TE}
,\tilde{g}_{2,6}^{R} =\frac{k_{2z}}{k_{s}^{2}}R_{2}^{TM}+\frac{k_{2}^{2}%
}{k_{s}^{2}k_{1z}}R_{2}^{TE},\nonumber\\
& \tilde{g}_{2,7}^{R} = \frac{k_{s}^{2}}{k_{1z}} R_{2}^{TM}, \tilde{g}%
_{2,8}^{R} = R_{2}^{TM}, \tilde{g}_{2,9}^{R} = \frac{k_{2z}}{k_{1z}}R_{2}%
^{TM},\nonumber\\
& g_{2,5}^{R} = 2\pi\int_{0}^{\infty} k_{s} \tilde{g}_{2,5}^{R} J_{0}%
(k_{s}\rho) e^{ik_{2z}z+ik_{1z}z^{\prime}+2ik_{2z}d} dk_{s}, g_{2,6}^{R} =
2\pi\int_{0}^{\infty} k_{s}^{3} \tilde{g}_{2,6}^{R} \frac{J_{2}(k_{s}\rho
)}{\rho^{2}}e^{ik_{2z}z+ik_{1z}z^{\prime}+2ik_{2z}d} dk_{s},\nonumber\\
& g_{2,7}^{R} = 2\pi\int_{0}^{\infty} k_{s} \tilde{g}_{2,7}^{R} J_{0}%
(k_{s}\rho) e^{ik_{2z}z+ik_{1z}z^{\prime}+2ik_{2z}d} dk_{s}, g_{2,8}^{R} =
2\pi\int_{0}^{\infty}k^{2}_{s} \tilde{g}_{2,8}^{R} \frac{J_{1}(k_{s}\rho
)}{\rho} e^{ik_{2z}z+ik_{1z}z^{\prime}+2ik_{2z}d} dk_{s},\nonumber\\
& g_{2,9}^{R} = 2\pi\int_{0}^{\infty}k^{2}_{s} \tilde{g}_{2,9}^{R} \frac
{J_{1}(k_{s}\rho)}{\rho} e^{ik_{2z}z+ik_{1z}z^{\prime}+2ik_{2z}d} dk_{s}.%
\end{align}
\newline
\noindent$\bullet$ In the third layer, again the spectral Green's function
have the same form as any transmitted fields in the second layer. Thus, Green's function can be expressed as
\begin{align}
G_{3xx}^{T}  & = \frac{1}{2}g_{3,5}^{T} - (\frac{1}{2}\rho^{2}-(y-y^{\prime
})^{2}) g_{3,6}^{T},\\
G_{3yy}^{T}  & = \frac{1}{2}g_{3,5}^{T} + (\frac{1}{2}\rho^{2}-(y-y^{\prime
})^{2}) g_{3,6}^{T},\\
G_{3zz}^{T}  & = g_{3,7}^{T},\\
G_{3xy}^{T}  & = G_{3xy}^{T} = -(x-x^{\prime})(y-y^{\prime})g_{3,6}^{T},\\
G_{3xz}^{T}  & =i(x-x^{\prime})g_{3,8}^{T},~~~ G_{3yz}^{T} = i(y-y^{\prime
})g_{3,8}^{T},\\
G_{3zx}^{T}  & = i(x-x^{\prime}) g_{3,9}^{T},~~~ G_{3zy}^{T} = i(y-y^{\prime})
g_{3,9}^{T},
\end{align}
where \begin{figure}[t]
\centering  \includegraphics[width=4.3in]{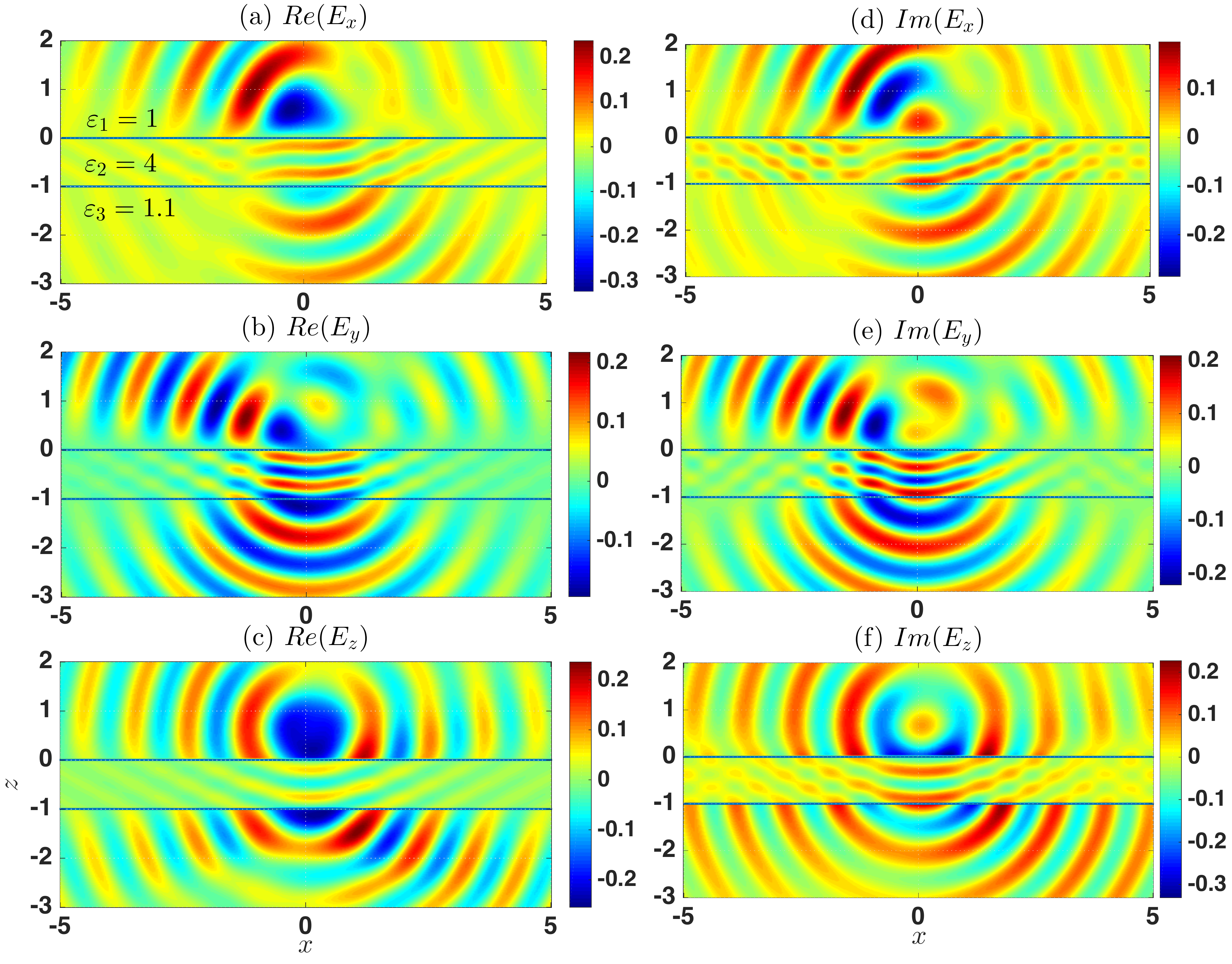}
\caption{Electric fields in a three-layer structure with layer interface at $z
= 0$ and $z = -1$. A dipole source is placed at $\mathbf{r}^{\prime}= (0.1,
-0.2, 0.5)$ and oriented along $\boldsymbol{\alpha}^{\prime}= (1/2, 1/2,
1/\sqrt{2})$ and fields are computed for $-5 \leq x \leq5$ and $-3 \leq z
\leq3$ for a fixed $y = 1.0$ with $\varepsilon_{1} = 1$, $\varepsilon_{2} =
4$, $\varepsilon_{3} = 1.1$, and $\lambda= 1$.}%
\label{three_layer_field}%
\end{figure}%
\begin{align}
\tilde{g}_{3,5}^{T}  & = k_{3z} A_{3}^{TM}+\frac{k_{3}^{2}}{k_{1z}}\frac
{\mu_{1}}{\mu_{3}} A_{3}^{TE}, \tilde{g}_{3,6}^{T}= \frac{k_{3z} }{k_{s}^{2}%
}A_{3}^{TM} - \frac{k_{3}^{2}}{k_{1z}k_{s}^{2}}\frac{\mu_{1}}{\mu_{3}}
A_{3}^{TE},\nonumber\\
\tilde{g}_{3,7}^{T}  & = \frac{k_{s}^{2}}{k_{1z}} A_{3}^{TM},\tilde{g}%
_{3,8}^{T} = \frac{k_{3z} A_{3}^{TM}}{k_{1z}} , \tilde{g}_{3,9}^{T} =
A_{3}^{TM},\nonumber\\
g_{3,5}^{T}  & = 2\pi\int_{0}^{\infty} k_{s} \tilde{g}_{3,5}^{T}J_{0}%
(k_{s}\rho) e^{-ik_{3z}z+ik_{1z}z^{\prime}}dk_{s}, g_{3,6}^{T} = 2\pi\int
_{0}^{\infty} k_{s}^{3} \tilde{g}_{3,6} \frac{J_{2}(k_{s}\rho)}{\rho^{2}}^{T}
e^{-ik_{3z}z+ik_{1z}z^{\prime}}dk_{s},\nonumber\\
g_{3,7}^{T}  & = 2\pi\int_{0}^{\infty} k_{s} \tilde{g}_{3,7}^{T} J_{0}%
(k_{s}\rho) e^{-ik_{3z}z+ik_{1z}z^{\prime}} dk_{s}, g_{3,8}^{T} = 2\pi\int
_{0}^{\infty}k^{2}_{s} \tilde{g}_{3,8}^{T} \frac{J_{1}(k_{s}\rho)}{\rho}
e^{-ik_{3z}z+ik_{1z}z^{\prime}} dk_{s},\nonumber\\
g_{3,9}^{T}  & = 2\pi\int_{0}^{\infty}k^{2}_{s} \tilde{g}_{3,9}^{T}
\frac{J_{1}(k_{s}\rho)}{\rho} e^{-ik_{3z}z+ik_{1z}z^{\prime}} dk_{s}.%
\end{align}

\subsection{Numerical results}
\begin{figure}[t]
\centering  \includegraphics[width=5in]{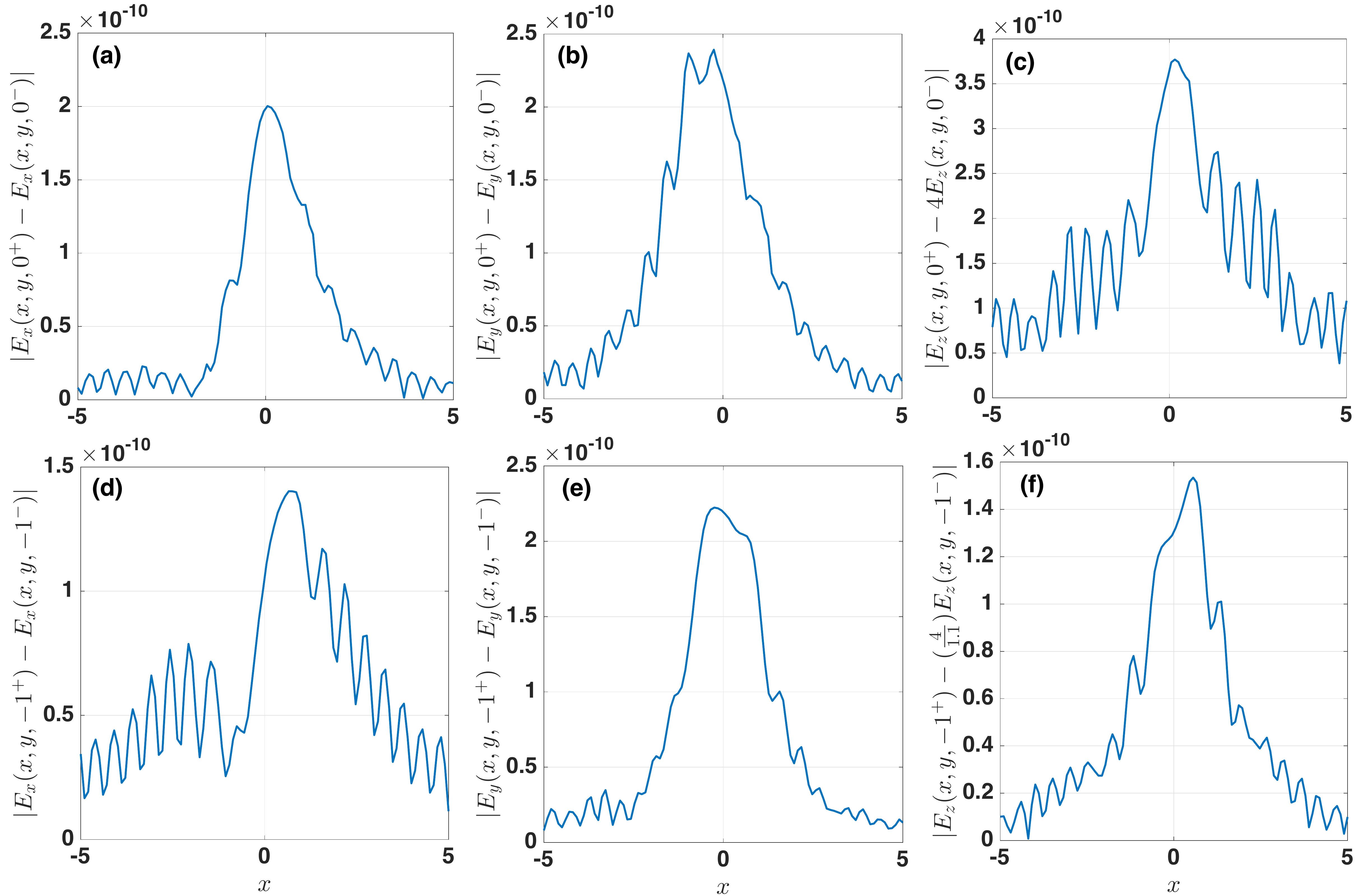}
\caption{Continuity of electric fields at both interfaces at $z = 0$ and $z =
-1$. (a) $|E_{x}(x,y,0^{+}) -E_{x}(x,y,0^{-})|$, (b) $|E_{y}(x,y,0^{+})
-E_{y}(x,y,0^{-})|$, (c) $|E_{z}(x,y,0^{+}) - 4E_{z}(x,y,0^{-})|$, (d)
$|E_{x}(x,y,-1^{+}) -E_{x}(x,y,-1^{-})|$, (e) $|E_{y}(x,y,-1^{+})
-E_{y}(x,y,-1^{-})|$, (f) $|E_{z}(x,y,-1^{+}) - \frac{1.1}{4}E_{z}%
(x,y,-1^{-})|$. A dipole source is placed at $\mathbf{r}^{\prime}= (0.1, -0.2,
0.5)$ and oriented along $\boldsymbol{\alpha}^{\prime}= (1/2, 1/2, 1/\sqrt
{2})$ and fields are computed for $-5 \leq x \leq5$ for a fixed $y = 1.0$ at
the layer interface $z = 0$ and $z = -1$ and with $\varepsilon_{1} = 1$,
$\varepsilon_{2} = 4$, $\varepsilon_{3} = 1.1$, and $\lambda= 1$.}%
\label{three_layer_error}%
\end{figure}

A three-layer structure is considered by placing two interfaces at $z = 0$ and $z = -1$. The relative permittivity is assigned as $\varepsilon_{1} = 1$, $\varepsilon_{2} = 4$, $\varepsilon_{3}= 1.1$ in each layer. The relative permeability $\{ \mu_{i} \}_{i = 1}^{3}$ is assumed to be $1$ in all layers. The wavelength $\lambda$ is set to be 1. The electric field is computed when a source is placed on top of the layered media at $\mathbf{r}^{\prime}= (0.1, -0.2, 0.5)$ and oriented along $\boldsymbol{\hat{\alpha}}^{\prime}= (1/2, 1/2, 1/\sqrt{2})$.  In Fig. \ref{three_layer_field}, all the components of electric field are plotted over $-5 \leq x \leq5$ and $-3 \leq z \leq3$ for a fixed $y = 1.0$. The continuity of the fields are checked at both interfaces $z = 0$ and $z = -1$ in Fig. \ref{three_layer_error} as accuracy of the Green's function. In all components, approximately $10^{-10}$ absolute error is obtained.

\subsection{Source in the second layer}

When a dipole source is placed in the second layer, the formula derived in the
previous subsections must be modified to accommodate multiple reflection and
transmission from the first and second interfaces. In the following, the electric field in each layer in the spectral domain are provided. Then, taking Sommerfeld integrals derives the electric field in the spatial domain. That will complete the derivation of Dyadic Green's function for a three-layer structure.

\subsubsection{Green's function in the spectral domain}

Let a dipole source is located in the second layer, then in the first and
third layer, there are only transmitted fields. However, in the second layer,
there are primary field, reflected fields from the bottom interface and the top
interface. The reflected fields from the bottom interface and top interface are
an up-going and a down-going waves, respectively. Thus, in each layer, the
$z$-components of field can be represented by
\begin{align}
\tilde{E}_{1z}  & = \tilde{E}_{1z}^{T}, \tilde{E}_{2z} = \tilde{E}_{2z}%
^{P}+\tilde{E}_{2z}^{U}+\tilde{E}_{2z}^{D}, \tilde{E}_{3z} = \tilde{E}%
_{3z}^{T},\\
\tilde{H}_{1z}  & = \tilde{H}_{1z}^{T}, \tilde{H}_{2z} = \tilde{H}_{2z}%
^{P}+\tilde{H}_{2z}^{U}+\tilde{H}_{2z}^{D}, \tilde{E}_{3z} = \tilde{H}%
_{3z}^{T}.
\end{align}
The primary field $\tilde{E}_{2z}^{P}$ is the same as the primary field in the free space. The field in the second layer can be expressed using new reflection coefficients $U^{TM, TE}$ and $D^{TM, TE}$ that represent the
amplitude of the up- and down-going waves, respectively.
\begin{align}
\tilde{E}_{2z}^{U}  & = \left(  \hat{z} \cdot\hat{\boldsymbol{\alpha}}%
^{\prime\prime}+\frac{1}{k_{2}^{2}} \partial_{z} \nabla\cdot\hat
{\boldsymbol{\alpha}}^{\prime\prime}\right)  \tilde{g}_{2, TM}^{U},~ \tilde
{H}_{2z}^{U} = -\frac{1}{i\omega\mu_{0}\mu_{2}}\hat{z} \cdot\nabla_{s}
\times\hat{\boldsymbol{\alpha}}^{\prime\prime}\tilde{g}_{2,TE}^{U}%
,\label{e2z_u}\\
\tilde{E}_{2z}^{D}  & = \left(  \hat{z} \cdot\hat{\boldsymbol{\alpha}}%
^{\prime\prime}+\frac{1}{k_{2}^{2}} \partial_{z} \nabla\cdot\hat
{\boldsymbol{\alpha}}^{\prime\prime}\right)  \tilde{g}_{2, TM}^{D},~ \tilde
{H}_{2z}^{D} = -\frac{1}{i\omega\mu_{0}\mu_{2}}\hat{z} \cdot\nabla_{s}
\times\hat{\boldsymbol{\alpha}}^{\prime\prime}\tilde{g}_{2,TE}^{D}
,\label{e2z_d}%
\end{align}
where
\begin{align}
\hat{\boldsymbol{\alpha}}^{\prime\prime} & = (-\alpha_{x}^{\prime},
-\alpha_{y}^{\prime}, \alpha_{z}^{\prime}),\\
\tilde{g}_{2,TM, TE}^{U}  &  = -\frac{\omega\mu_{0} \mu_{2}}{8 \pi^{2}}
U^{TM,TE}\frac{ e^{ik_{x}(x-x^{\prime})+ik_{y}(y-y^{\prime})}}{k_{2z}}
e^{ik_{2z}z},\\
\tilde{g}_{2,TM, TE}^{D}  &  = -\frac{\omega\mu_{0} \mu_{2}}{8 \pi^{2}}
D^{TM, TE}\frac{ e^{ik_{x}(x-x^{\prime})+ik_{y}(y-y^{\prime})}}{k_{2z}}
e^{-ik_{2z}z}.
\end{align}
In the first and third layer, the transmitted parts are given by
\begin{align}
\tilde{E}_{1z}^{T}  & = \left(  \hat{z} \cdot\hat{\boldsymbol{\alpha}}%
_{1}^{\prime\prime\prime}+\frac{1}{k_{1}^{2}} \partial_{z} \nabla\cdot
\hat{\boldsymbol{\alpha}}_{1}^{\prime\prime\prime}\right)  \tilde{g}_{1,
TM}^{T},~ \tilde{H}_{1z}^{T} = \frac{1}{i\omega\mu_{0}\mu_{1}} \frac{k_{1z}%
}{k_{2z}} \hat{z} \cdot\nabla_{s} \times\hat{\boldsymbol{\alpha}}_{1}%
^{\prime\prime\prime}\tilde{g}_{1,TE}^{T},\label{e1z_t}\\
\tilde{E}_{3z}^{T}  & = \left(  \hat{z} \cdot\hat{\boldsymbol{\alpha}}%
_{3}^{\prime\prime\prime}+\frac{1}{k_{3}^{2}} \partial_{z} \nabla\cdot
\hat{\boldsymbol{\alpha}}_{3}^{\prime\prime\prime}\right)  \tilde{g}_{3,
TM}^{T},~ \tilde{H}_{3z}^{T} = \frac{1}{i\omega\mu_{0}\mu_{3}} \frac{k_{3z}%
}{k_{2z}} \hat{z} \cdot\nabla_{s} \times\hat{\boldsymbol{\alpha}}_{3}%
^{\prime\prime\prime}\tilde{g}_{3,TE}^{T},\label{e3z_t}%
\end{align}
where
\begin{align}
& \hat{\boldsymbol{\alpha}}_{1}^{\prime\prime\prime}= (\frac{k_{2z}}{k_{1z}%
}\alpha_{x}^{\prime}, \frac{k_{2z}}{k_{1z}}\alpha_{y}^{\prime}, \alpha
_{z}^{\prime}), \hat{\boldsymbol{\alpha}}_{3}^{\prime\prime\prime}=
(\frac{k_{2z}}{k_{3z}}\alpha_{x}^{\prime}, \frac{k_{2z}}{k_{3z}}\alpha
_{y}^{\prime}, \alpha_{z}^{\prime}),\\
& \tilde{g}_{1,TM, TE}^{T} = -\frac{\omega\mu_{0} \mu_{1}}{8 \pi^{2}}
A_{1}^{TM,TE}\frac{ e^{ik_{x}(x-x^{\prime})+ik_{y}(y-y^{\prime})}}{k_{2z}}
e^{ik_{1z}z},\\
& \tilde{g}_{3,TM, TE}^{T} = -\frac{\omega\mu_{0} \mu_{3}}{8 \pi^{2}}
A_{3}^{TM,TE}\frac{ e^{ik_{x}(x-x^{\prime})+ik_{y}(y-y^{\prime})}}{k_{2z}}
e^{-ik_{3z}z}.
\end{align}
In the above, $A_{1}^{TM, TE}$, $A_{3}^{TM, TE}$, $D^{TM, TE}$, and $U^{TM,
TE}$ (See the Ref. \cite{chewbook} for their derivation) are given by
\begin{align}
A_{1}^{TM, TE}  &  = \frac{T_{21}^{TM, TE} \left( e^{-ik_{2z}z^{\prime}}\pm
R_{23}^{TM, TE} e^{ik_{2z}z^{\prime}+2ik_{2z}d} \right) }{1-R_{23}^{TM,
TE}R_{21}^{TM, TE} e^{2ik_{2z}d}},\\
D^{TM, TE}  &  = \frac{R_{21}^{TM, TE} \left( e^{-ik_{2z}z^{\prime}}\pm
R_{23}^{TM, TE} e^{ik_{2z}z^{\prime}+2ik_{2z}d} \right) }{1-R_{23}^{TM,
TE}R_{21}^{TM, TE} e^{2ik_{2z}d}},\\
U^{TM, TE}  &  = \frac{R_{23}^{TM, TE} e^{2ik_{2z}d} \left( e^{ik_{2z}%
z^{\prime}}\pm R_{21}^{TM, TE} e^{-ik_{2z}z^{\prime}} \right) }{1-R_{23}^{TM,
TE}R_{21}^{TM, TE} e^{2ik_{2z}d}},\\
A_{3}^{TM, TE}  &  = \frac{T_{23}^{TM, TE} e^{i\left( k_{2z}-k_{3z}\right) d}
\left( e^{ik_{2z}z^{\prime}}\pm R_{21}^{TM, TE} e^{-ik_{2z}z^{\prime}} \right)
}{1-R_{23}^{TM, TE}R_{21}^{TM, TE} e^{2ik_{2z}d}}.
\end{align}
In each layer, again the transverse component must be derived using Maxwell's equations. The final simplified formula is presented in the following:\newline

\noindent$\bullet$ In the first layer, the electric field in the spectral domain
is
\begin{align}
\left[
\begin{array}
[c]{c}%
\tilde{E}_{1x}\\
\tilde{E}_{1y}\\
\tilde{E}_{1z}%
\end{array}
\right]   & = \left[
\begin{array}
[c]{c}%
\tilde{E}_{1x}^{T}\\
\tilde{E}_{1y}^{T}\\
\tilde{E}_{1z}^{T}%
\end{array}
\right]  =-\frac{1}{8\pi^{2}\omega\varepsilon_{0}\varepsilon_{1}}%
\tilde{\mathbf{G}}_{1}^{T} \alpha^{\prime}= -\frac{1}{8\pi^{2}\omega
\varepsilon_{0}\varepsilon_{1}} \left[
\begin{array}
[c]{ccc}%
\tilde{G}^{T}_{1xx} & \tilde{G}^{T}_{1xy} & \tilde{G}^{T}_{1xz}\\
\tilde{G}^{T}_{1yx} & \tilde{G}^{T}_{1yy} & \tilde{G}^{T}_{1yz}\\
\tilde{G}^{T}_{1zx} & \tilde{G}^{T}_{1zy} & \tilde{G}^{T}_{1zz}%
\end{array}
\right]  \left[
\begin{array}
[c]{c}%
\alpha_{x}^{\prime}\\
\alpha_{y}^{\prime}\\
\alpha_{z}^{\prime}%
\end{array}
\right] ,
\end{align}
where
\begin{align}
\tilde{G}_{1xx}^{T}  & =\left(  -\partial_{x}^{2} \frac{k_{1z}}{k_{s}^{2}}
A_{1}^{TM} -\partial_{y}^{2} \frac{k_{1}^{2}}{k_{s}^{2} k_{2z}}A_{1}%
^{TE}\right) e^{ik_{x}(x-x^{\prime})+ik_{y}(y-y^{\prime})} e^{ik_{1z}z},\\
\tilde{G}_{1yy}^{T}  & =\left(  -\partial_{y}^{2} \frac{k_{1z}}{k_{s}^{2}%
}A_{1}^{TM}-\partial_{x}^{2} \frac{k_{1}^{2}}{k_{s}^{2} k_{2z}} A_{1}^{TE}
\right) e^{ik_{x}(x-x^{\prime})+ik_{y}(y-y^{\prime})} e^{ik_{1z}z},\\
\tilde{G}_{1zz}^{T}  & =\left( \frac{k_{s}^{2} }{k_{2z}}A_{1}^{TM} \right)
e^{ik_{x}(x-x^{\prime})+ik_{y}(y-y^{\prime})} e^{ik_{1z}z},\\
\tilde{G}_{1xy}^{T}  & =\tilde{G}_{yx}^{T} = \left(  -\partial_{x}
\partial_{y} \frac{k_{1z}}{k_{s}^{2}}A_{1}^{TM}+\partial_{x}\partial_{y}
\frac{k_{1}^{2}}{k_{s}^{2} k_{2z}} A_{1}^{TE} \right) e^{ik_{x}(x-x^{\prime
})+ik_{y}(y-y^{\prime})} e^{ik_{1z}z},\\
\tilde{G}_{1xz}^{T}  & = \left(  \partial_{x} \partial_{z} \frac{A_{1}^{TM}%
}{k_{2z}} \right)  e^{ik_{x}(x-x^{\prime})+ik_{y}(y-y^{\prime})} e^{ik_{1z}z},
\tilde{G}_{1yz}^{T} = \left(  \partial_{y} \partial_{z} \frac{A_{1}^{TM}%
}{k_{2z}} \right) e^{ik_{x}(x-x^{\prime})+ik_{y}(y-y^{\prime})} e^{ik_{1z}%
z},\\
\tilde{G}_{1zx} ^{T}  & = \left( \partial_{x} \partial_{z} \frac{A_{1}^{TM}%
}{k_{1z}} \right) e^{ik_{x}(x-x^{\prime})+ik_{y}(y-y^{\prime})} e^{ik_{1z}z},
\tilde{G}_{1zy} ^{T} = \left( \partial_{y} \partial_{z} \frac{A_{1}^{TM}%
}{k_{1z}} \right) e^{ik_{x}(x-x^{\prime})+ik_{y}(y-y^{\prime})} e^{ik_{1z}z}.
\end{align}
\newline

\noindent$\bullet$ In the third layer, the same calculation applies and the
electric field in the spectral domain is
\begin{align}
\left[
\begin{array}
[c]{c}%
\tilde{E}_{3x}\\
\tilde{E}_{3y}\\
\tilde{E}_{3z}%
\end{array}
\right]   & = \left[
\begin{array}
[c]{c}%
\tilde{E}_{3x}^{T}\\
\tilde{E}_{3y}^{T}\\
\tilde{E}_{3z}^{T}%
\end{array}
\right]  =-\frac{1}{8\pi^{2}\omega\varepsilon_{0}\varepsilon_{3}}%
\tilde{\mathbf{G}}_{3}^{T} \alpha^{\prime}= -\frac{1}{8\pi^{2}\omega
\varepsilon_{0}\varepsilon_{3}} \left[
\begin{array}
[c]{ccc}%
\tilde{G}^{T}_{3xx} & \tilde{G}^{T}_{3xy} & \tilde{G}^{T}_{3xz}\\
\tilde{G}^{T}_{3yx} & \tilde{G}^{T}_{3yy} & \tilde{G}^{T}_{3yz}\\
\tilde{G}^{T}_{3zx} & \tilde{G}^{T}_{3zy} & \tilde{G}^{T}_{3zz}%
\end{array}
\right]  \left[
\begin{array}
[c]{c}%
\alpha_{x}^{\prime}\\
\alpha_{y}^{\prime}\\
\alpha_{z}^{\prime}%
\end{array}
\right] ,
\end{align}
where
\begin{align}
\tilde{G}_{3xx}^{T}  & =\left(  -\partial_{x}^{2} \frac{k_{3z}}{k_{s}^{2}}
A_{3}^{TM} -\partial_{y}^{2} \frac{k_{3}^{2}}{k_{s}^{2} k_{2z}}A_{3}%
^{TE}\right) e^{ik_{x}(x-x^{\prime})+ik_{y}(y-y^{\prime})} e^{-ik_{3z}z},\\
\tilde{G}_{3yy}^{T}  & =\left(  -\partial_{y}^{2} \frac{k_{3z}}{k_{s}^{2}%
}A_{3}^{TM}-\partial_{x}^{2} \frac{k_{3}^{2}}{k_{s}^{2} k_{2z}} A_{3}^{TE}
\right) e^{ik_{x}(x-x^{\prime})+ik_{y}(y-y^{\prime})} e^{-ik_{3z}z},\\
\tilde{G}_{3zz}^{T}  & =\left( \frac{k_{s}^{2} A_{3}^{TM}}{k_{2z}} \right)
e^{ik_{x}(x-x^{\prime})+ik_{y}(y-y^{\prime})} e^{-ik_{3z}z},\\
\tilde{G}_{3xy}^{T}  & =\tilde{G}_{yx}^{T} = \left(  -\partial_{x}
\partial_{y} \frac{k_{3z}}{k_{s}^{2}}A_{3}^{TM}+\partial_{x}\partial_{y}
\frac{k_{3}^{2}}{k_{s}^{2} k_{2z}} A_{3}^{TE} \right) e^{ik_{x}(x-x^{\prime
})+ik_{y}(y-y^{\prime})} e^{-ik_{3z}z},\\
\tilde{G}_{3xz}^{T}  & = \left(  \partial_{x} \partial_{z} \frac{A_{3}^{TM}%
}{k_{2z}} \right)  e^{ik_{x}(x-x^{\prime})+ik_{y}(y-y^{\prime})} e^{-ik_{3z}%
z}, \tilde{G}_{3yz}^{T} = \left(  \partial_{y} \partial_{z} \frac{A_{3}^{TM}%
}{k_{2z}} \right) e^{ik_{x}(x-x^{\prime})+ik_{y}(y-y^{\prime})} e^{-ik_{3z}%
z},\\
\tilde{G}_{3zx} ^{T}  & = \left( \partial_{x} \partial_{z} \frac{A_{3}^{TM}%
}{k_{3z}} \right) e^{ik_{x}(x-x^{\prime})+ik_{y}(y-y^{\prime})} e^{-ik_{3z}z},
\tilde{G}_{3zy} ^{T} = \left( \partial_{y} \partial_{z} \frac{A_{3}^{TM}%
}{k_{3z}} \right) e^{ik_{x}(x-x^{\prime})+ik_{y}(y-y^{\prime})} e^{-ik_{3z}z}.
\end{align}
\newline

\noindent$\bullet$ In the second layer, the electric field has three parts that can
be expressed using the Green's function notation. The derivation of the up-going wave Green's function ($\tilde{\mathbf{G}}_{2}^{U}$) and the down-going
wave Green's function ($\tilde{\mathbf{G}}_{2}^{D}$) are similar to that
of reflection fields in both two- and three-layer structures. Derivation are not
so difficult but needs some attention on $k_{2z}$ because there is $k_{2z}$ in
the denominator of $\tilde{g}_{2, TM,TE}^{U,D}$ and $\tilde{g}_{3, TM,TE}^{T}$
instead of $k_{1z}$ compared with the case when the source is in the first layer. In the following,
both the up- and down-going wave Green's functions are listed.%

\begin{align}
\left[
\begin{array}
[c]{c}%
\tilde{E}_{2x}\\
\tilde{E}_{2y}\\
\tilde{E}_{2z}%
\end{array}
\right]   & = \left[
\begin{array}
[c]{c}%
\tilde{E}_{2x}^{P}+\tilde{E}_{2x}^{D}+\tilde{E}_{2x}^{U}\\
\tilde{E}_{2y}^{P}+\tilde{E}_{2y}^{D}+\tilde{E}_{2y}^{U}\\
\tilde{E}_{2z}^{P}+\tilde{E}_{2z}^{D}+\tilde{E}_{2z}^{U}%
\end{array}
\right]  =\left[ \tilde{\mathbf{G}}_{2}^{P}-\frac{1}{8\pi^{2}\omega
\varepsilon_{0}\varepsilon_{2}}\left( \tilde{\mathbf{G}}_{2}^{D}%
+\tilde{\mathbf{G}}_{2}^{U} \right) \right] \boldsymbol{\hat{\alpha}}^{\prime}\nonumber\\
& =\left[ \tilde{\mathbf{G}}_{2}^{P} -\frac{1}{8\pi^{2}\omega\varepsilon
_{0}\varepsilon_{2}}\left(  \left[
\begin{array}
[c]{ccc}%
\tilde{G}^{D}_{2xx} & \tilde{G}^{D}_{2xy} & \tilde{G}^{D}_{2xz}\\
\tilde{G}^{D}_{2yx} & \tilde{G}^{D}_{2yy} & \tilde{G}^{D}_{2yz}\\
\tilde{G}^{D}_{2zx} & \tilde{G}^{D}_{2zy} & \tilde{G}^{D}_{2zz}%
\end{array}
\right]  +\left[
\begin{array}
[c]{ccc}%
\tilde{G}^{U}_{2xx} & \tilde{G}^{U}_{2xy} & \tilde{G}^{U}_{2xz}\\
\tilde{G}^{U}_{2yx} & \tilde{G}^{U}_{2yy} & \tilde{G}^{U}_{2yz}\\
\tilde{G}^{U}_{2zx} & \tilde{G}^{U}_{2zy} & \tilde{G}^{U}_{2zz}%
\end{array}
\right]  \right)  \right] \left[
\begin{array}
[c]{c}%
\alpha_{x}^{\prime}\\
\alpha_{y}^{\prime}\\
\alpha_{z}^{\prime}%
\end{array}
\right] ,
\end{align}
where
\begin{align}
\tilde{G}_{2xx}^{D}  &  = \left( \partial_{x}^{2} \frac{k_{2z}}{k_{s}^{2}}
D^{TM} -\partial_{y}^{2} \frac{k_{2}^{2}}{k_{s}^{2} k_{2z}}D^{TE}\right)
e^{ik_{x}(x-x^{\prime})+ik_{y}(y-y^{\prime})} e^{-ik_{2z}z},\\
\tilde{G}_{2yy}^{D}  &  = \left( \partial_{y}^{2} \frac{k_{2z}}{k_{s}^{2}}
D^{TM} -\partial_{x}^{2} \frac{k_{2}^{2}}{k_{s}^{2} k_{2z}}D^{TE}\right)
e^{ik_{x}(x-x^{\prime})+ik_{y}(y-y^{\prime})} e^{-ik_{2z}z},\\
\tilde{G}_{2zz}^{D}  &  = \left( \frac{k_{s}^{2} D^{TM}}{k_{2z}} \right)
e^{ik_{x}(x-x^{\prime})+ik_{y}(y-y^{\prime})} e^{-ik_{2z}z},\\
\tilde{G}_{2xy}^{D}  &  = \tilde{G}_{2yx}^{D} = \left( \partial_{x}%
\partial_{y} \frac{k_{2z}}{k_{s}^{2}} D^{TM} +\partial_{x}\partial_{y}
\frac{k_{2}^{2}}{k_{s}^{2} k_{2z}}D^{TE}\right)  e^{ik_{x}(x-x^{\prime
})+ik_{y}(y-y^{\prime})} e^{-ik_{2z}z},\\
\tilde{G}_{2xz}^{D}  &  = -\tilde{G}_{2zx}^{D} = \partial_{x}\partial_{z}
\frac{D^{TM}}{k_{2z}}e^{ik_{x}(x-x^{\prime})+ik_{y}(y-y^{\prime})}
e^{-ik_{2z}z},\\
\tilde{G}_{2yz}^{D}  &  = -\tilde{G}_{2zy}^{D} = \partial_{y}\partial_{z}
\frac{D^{TM}}{k_{2z}}e^{ik_{x}(x-x^{\prime})+ik_{y}(y-y^{\prime})}
e^{-ik_{2z}z},
\end{align}
and
\begin{align}
\tilde{G}_{2xx}^{U}  &  = \left( \partial_{x}^{2} \frac{k_{2z}}{k_{s}^{2}}
U^{TM} -\partial_{y}^{2} \frac{k_{2}^{2}}{k_{s}^{2} k_{2z}}U^{TE}\right)
e^{ik_{x}(x-x^{\prime})+ik_{y}(y-y^{\prime})} e^{ik_{2z}z},\\
\tilde{G}_{2yy}^{U}  &  = \left( \partial_{y}^{2} \frac{k_{2z}}{k_{s}^{2}}
U^{TM} -\partial_{x}^{2} \frac{k_{2}^{2}}{k_{s}^{2} k_{2z}}U^{TE}\right)
e^{ik_{x}(x-x^{\prime})+ik_{y}(y-y^{\prime})} e^{ik_{2z}z},\\
\tilde{G}_{2zz}^{U}  &  = \left( \frac{k_{s}^{2} U^{TM}}{k_{2z}} \right)
e^{ik_{x}(x-x^{\prime})+ik_{y}(y-y^{\prime})} e^{ik_{2z}z},\\
\tilde{G}_{2xy}^{U}  &  = \tilde{G}_{2yx}^{U} = \left( \partial_{x}%
\partial_{y} \frac{k_{2z}}{k_{s}^{2}} U^{TM} +\partial_{x}\partial_{y}
\frac{k_{2}^{2}}{k_{s}^{2} k_{2z}}U^{TE}\right)  e^{ik_{x}(x-x^{\prime
})+ik_{y}(y-y^{\prime})} e^{ik_{2z}z},\\
\tilde{G}_{2xz}^{U}  &  = -\tilde{G}_{2zx}^{U} = \partial_{x}\partial_{z}
\frac{U^{TM}}{k_{2z}}e^{ik_{x}(x-x^{\prime})+ik_{y}(y-y^{\prime})}
e^{ik_{2z}z},\\
\tilde{G}_{2yz}^{U}  &  = -\tilde{G}_{2zy}^{U} = \partial_{y}\partial_{z}
\frac{U^{TM}}{k_{2z}}e^{ik_{x}(x-x^{\prime})+ik_{y}(y-y^{\prime})}
e^{ik_{2z}z}.
\end{align}

\subsubsection{Green's function in the spatial domain}

As expected from the previous sections, the inverse Fourier transform are applied to the spectral Green's function to obtain the one in the spatial domain. Most of basic
computations are already performed while deriving the two- and three-layer Green's functions. Therefore, without any derivation, the Green's function in the spatial domain is presented below.\newline

\noindent$\bullet$ In the first layer,
\begin{align}
G_{1xx}^{T}  &  = \frac{1}{2} g_{1,5}^{T} - \left( \frac{1}{2}\rho^{2} -
(y-y^{\prime})^{2} \right)  g_{1,6}^{T},\\
G_{1yy}^{T}  &  = \frac{1}{2} g_{1,5}^{T} + \left( \frac{1}{2}\rho^{2} -
(y-y^{\prime})^{2} \right)  g_{1,6}^{T},\\
G_{1zz}^{T}  &  = g_{1,7}^{T},\\
G_{1xy}^{T}  &  = G_{1yx}^{T} = -(x-x^{\prime})(y-y^{\prime})g_{1,6}^{T},\\
G_{1xz}^{T}  &  = -i(x-x^{\prime})g_{1,8}^{T}, G_{1yz}^{T} = -i(y-y^{\prime
})g_{1,8}^{T},\\
G_{1zx}^{T}  &  = -i(x-x^{\prime})g_{1,9}^{T}, G_{1zy}^{T} = -i(y-y^{\prime
})g_{1,9}^{T},
\end{align}
where
\begin{align}
\tilde{g}_{1,5}^{T}  & = k_{1z}A_{1}^{TM}+\frac{k_{1}^{2}}{k_{2z}}A_{1}%
^{TE},\tilde{g}_{1,6}^{T} = \frac{k_{1z}A_{1}^{TM}}{k_{s}^{2}}-\frac{k_{1}%
^{2}}{k_{2z}k_{s}^{2}}A_{1}^{TE}, \tilde{g}_{1,7}^{T} = \frac{k_{s}^{2}%
}{k_{2z}}A_{1}^{TM}, \tilde{g}_{1,8}^{T} = \frac{k_{1z}}{k_{2z}} A_{1}^{TM},
\tilde{g}_{1,9}^{T} = A_{1}^{TM},\nonumber\\
g_{1,5}^{T}  & = 2\pi\int_{0}^{\infty} k_{s} \tilde{g}_{1,5}^{T} J_{0}%
(k_{s}\rho) e^{ik_{1z}z} dk_{s},g_{1,6}^{T} = 2\pi\int_{0}^{\infty} k^{3}_{s}
\tilde{g}_{1,6}^{T} \frac{J_{2}(k_{s}\rho) }{\rho^{2}}e^{ik_{1z}z}
dk_{s}\nonumber\\
g_{1,7}^{T}  & = 2\pi\int_{0}^{\infty} k_{s} \tilde{g}_{1,7}^{T}J_{0}%
(k_{s}\rho) e^{ik_{1z}z}dk_{s}, {g}_{1,8}^{T} = 2\pi\int_{0}^{\infty}
k_{s}^{2} \tilde{g}_{1,8}^{T} \frac{J_{1}(k_{s}\rho)}{\rho} e^{ik_{1z}z}
dk_{s},\nonumber\\
g_{1,9}^{T}  & =2\pi\int_{0}^{\infty} k^{2}_{s} \tilde{g}_{1,9}^{T}
\frac{J_{1}(k_{s}\rho)}{\rho} e^{ik_{1z}z} dk_{s}.
\end{align}
\newline
\noindent$\bullet$ In the third layer, all the formulas take almost same form
as the first layer except the direction of the field. Therefore, they are given by
\begin{align}
G_{3xx}^{T}  & = \frac{1}{2}g_{3,5}^{T} - \left( \frac{1}{2}\rho^{2} -
(y-y^{\prime})^{2} \right)  g_{3,6}^{T},\\
G_{3yy}^{T}  & = \frac{1}{2} g_{3,5}^{T} + \left( \frac{1}{2}\rho^{2} -
(y-y^{\prime})^{2} \right)  g_{3,6}^{T},\\
G_{3zz}^{T}  & = g_{3,7}^{T},\\
G_{3xy}^{T}  & = G_{3yx}^{T} = -(x-x^{\prime})(y-y^{\prime}) g_{3,6}^{T},\\
G_{3xz}^{T}  &  = i(x-x^{\prime})g_{3,8}^{T}, G_{3yz}^{T} = i(y-y^{\prime
})g_{3,8}^{T},\\
G_{3zx}^{T}  &  = i(x-x^{\prime})g_{3,9}^{T}, G_{3zy}^{T} = i(y-y^{\prime
})g_{3,9}^{T},
\end{align}
where
\begin{align}
\tilde{g}_{3,5}^{T}  & = k_{3z}A_{3}^{TM}+\frac{k_{3}^{2}}{k_{2z}}A_{3}%
^{TE},\tilde{g}_{3,6}^{T} = \frac{k_{3z}A_{3}^{TM}}{k_{s}^{2}}-\frac{k_{3}%
^{2}}{k_{2z}k_{s}^{2}}A_{3}^{TE}, \tilde{g}_{3,7}^{T} = \frac{k_{s}^{2}%
}{k_{2z}}A_{3}^{TM},\tilde{g}_{3,8}^{T} = \frac{k_{3z}}{k_{2z}} A_{3}^{TM},
\tilde{g}_{3,9}^{T} = A_{3}^{TM},\nonumber\\
g_{3,5}^{T}  & = 2\pi\int_{0}^{\infty} k_{s} \tilde{g}_{3,5}^{T} J_{0}%
(k_{s}\rho) e^{-ik_{3z}z} dk_{s},g_{3,6}^{T} = 2\pi\int_{0}^{\infty} k^{3}_{s}
\tilde{g}_{3,6}^{T} \frac{J_{2}(k_{s}\rho) }{\rho^{2}}e^{-ik_{3z}z}
dk_{s},\nonumber\\
g_{3,7}^{T}  & = 2\pi\int_{0}^{\infty} k_{s} \tilde{g}_{3,7}^{T}J_{0}%
(k_{s}\rho) e^{-ik_{3z}z}dk_{s}, {g}_{3,8}^{T} = 2\pi\int_{0}^{\infty}
k_{s}^{2} \tilde{g}_{3,8}^{T} \frac{J_{1}(k_{s}\rho)}{\rho} e^{-ik_{3z}z}
dk_{s},\nonumber\\
g_{3,9}^{T}  & =2\pi\int_{0}^{\infty} k^{2}_{s} \tilde{g}_{3,9}^{T}
\frac{J_{1}(k_{s}\rho)}{\rho} e^{-ik_{3z}z} dk_{s}.
\end{align}
\newline
\noindent$\bullet$ In the second layer, both the up- and down-going waves are
reflected wave from the interface. Thus, the Green's function follows similar
formula as the reflected field in both two- and three-layer structures.
However, again one must be careful about the sign because of direction. The
up-going wave Green's function is obtained as
\begin{align}
G_{2xx}^{U}  & = -\frac{1}{2}g_{2,5}^{U}+\left( \frac{1}{2}\rho^{2}%
-(y-y^{\prime})^{2} \right)  g_{2,6}^{U},\\
G_{2yy}^{U}  & = -\frac{1}{2}g_{2,5}^{U}-\left( \frac{1}{2}\rho^{2}%
-(y-y^{\prime})^{2} \right)  g_{2,6}^{U},\\
G_{2zz}^{U}  & = g_{2,7}^{U},\\
G_{2xy}^{U}  &  = G_{2yx}^{U} = (x-x^{\prime})(y-y^{\prime})g_{2,6}^{U},\\
G_{2xz}^{U}  &  = -G_{2zx}^{U} = -i(x-x^{\prime})g_{2,8}^{U},\\
G_{2yz}^{U}  &  = -G_{2zy}^{U} = -i(y-y^{\prime})g_{2,8}^{U},
\end{align}
where
\begin{align}
\tilde{g}_{2,5}^{U}  & = k_{2z}U^{TM}-\frac{k_{2}^{2}}{k_{2z}}U^{TE},
\tilde{g}_{2,6}^{U} = \frac{k_{2z}U^{TM}}{k_{s}^{2}}+\frac{k_{2}^{2}}%
{k_{s}^{2}k_{2z}}U^{TE},\tilde{g}_{2,7}^{U} = \frac{k_{s}^{2}}{k_{2z}}%
U^{TM},\tilde{g}_{2,8}^{U} = U^{TM},\nonumber\\
g_{2,5}^{U}  & = 2\pi\int_{0}^{\infty} k_{s} \tilde{g}_{2,5}^{U} J_{0}%
(k_{s}\rho)e^{ik_{2z}z} dk_{s}, g_{2,6}^{U} = 2\pi\int_{0}^{\infty} k^{3}_{s}
\tilde{g}_{2,6}^{U} \frac{J_{2}(k_{s}\rho)}{\rho^{2}}e^{ik_{2z}z} dk_{s},\nonumber\\
g_{2,7}^{U}  & = 2\pi\int_{0}^{\infty} k_{s} \tilde{g}_{2,7}^{U} J_{0}%
(k_{s}\rho) e^{ik_{2z}z} dk_{s}, g_{2,8}^{U} = 2\pi\int_{0}^{\infty} k_{s}^{2}
\tilde{g}_{2,8}^{U} \frac{J_{1}(k_{s}\rho)}{\rho}e^{ik_{2z}z} dk_{s}.
\end{align}
The down-going wave Green's function is given by
\begin{align}
G_{2xx}^{D} & = -\frac{1}{2}g_{2,5}^{D}+\left( \frac{1}{2}\rho^{2}%
-(y-y^{\prime})^{2} \right)  g_{2,6}^{D},\\
G_{2yy}^{D}  & = -\frac{1}{2}g_{2,5}^{D}-\left( \frac{1}{2}\rho^{2}%
-(y-y^{\prime})^{2} \right)  g_{2,6}^{D},\\
G_{2zz}^{D}  & = g_{2,7}^{D},\\
G_{2xy}^{D}  & = G_{2yx}^{D} =(x-x^{\prime})(y-y^{\prime}) g_{2,6}^{D},\\
G_{2xz}^{D}  & = -G_{2zx}^{D} = i(x-x^{\prime}) g_{2,8}^{D},\\
G_{2yz}^{D}  & = -G_{2zy}^{D} = i(y-y^{\prime}) g_{2,8}^{D},
\end{align}
where
\begin{align}
\tilde{g}_{2,5}^{D}  & = k_{2z}D^{TM}-\frac{k_{2}^{2}}{k_{2z}}D^{TE},
\tilde{g}_{2,6}^{D} = \frac{k_{2z}D^{TM}}{k_{s}^{2}}+\frac{k_{2}^{2}}%
{k_{s}^{2}k_{2z}}D^{TE}, \tilde{g}_{2,7}^{D} = \frac{k_{s}^{2}}{k_{2z}}%
D^{TM},\tilde{g}_{2,8}^{D} = D^{TM},\nonumber\\
g_{2,5}^{D}  & = 2\pi\int_{0}^{\infty} k_{s} \tilde{g}_{2,5}^{D} J_{0}%
(k_{s}\rho)e^{-ik_{2z}z} dk_{s}, g_{2,6}^{D} = 2\pi\int_{0}^{\infty} k^{3}_{s}
\tilde{g}_{2,6}^{D} \frac{J_{2}(k_{s}\rho)}{\rho^{2}}e^{-ik_{2z}z} dk_{s},\nonumber\\
g_{2,7}^{D}  & = 2\pi\int_{0}^{\infty} k_{s} \tilde{g}_{2,7}^{D} J_{0}%
(k_{s}\rho) e^{-ik_{2z}z} dk_{s} , g_{2,8}^{D} = 2\pi\int_{0}^{\infty}
k_{s}^{2} \tilde{g}_{2,8}^{D} \frac{J_{1}(k_{s}\rho)}{\rho}e^{-ik_{2z}z}
dk_{s}.
\end{align}
\begin{figure}[t]
\centering  \includegraphics[width=4.4in]{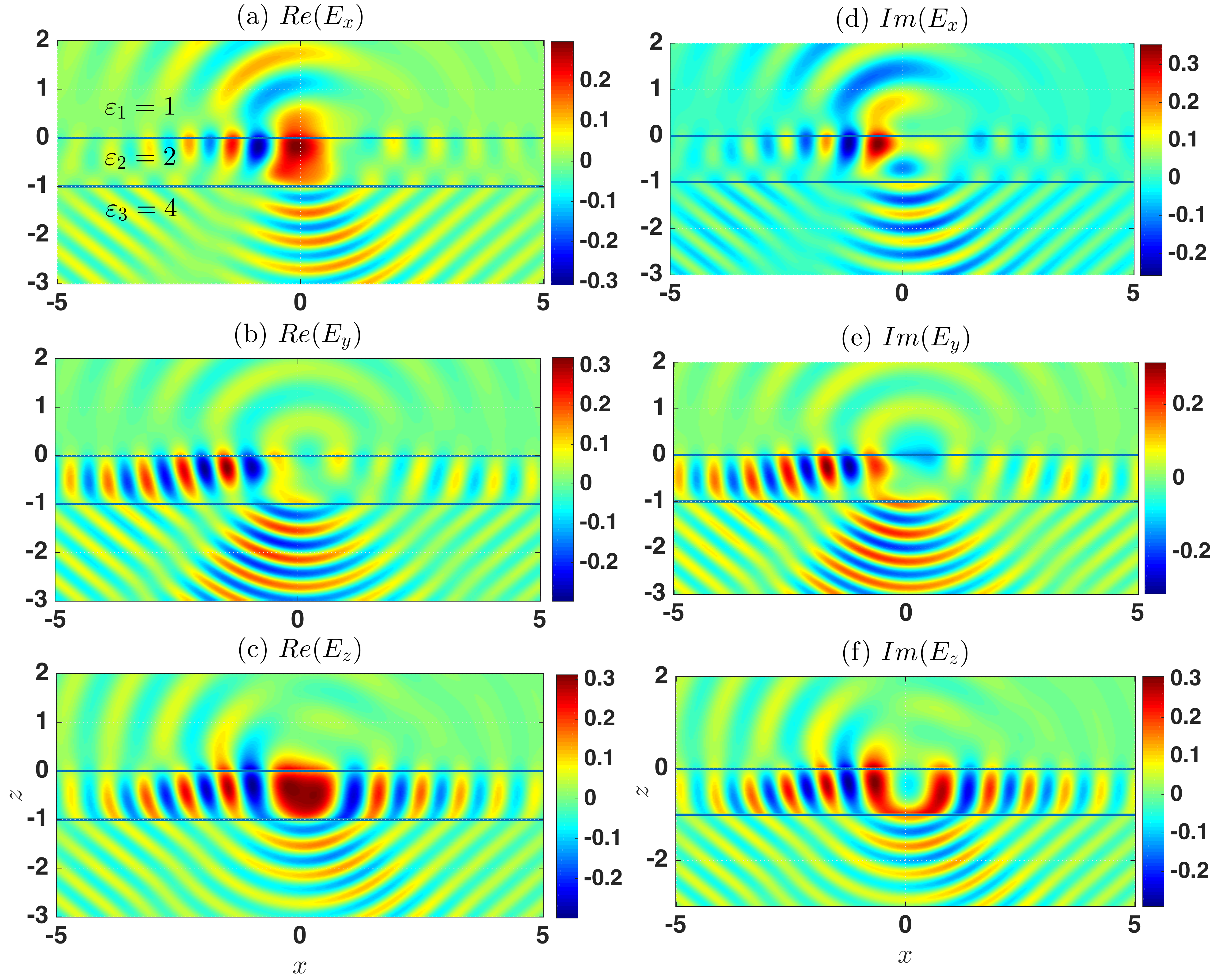}
\caption{Electric fields in a three-layer structure with layer interface at $z
= 0$ and $z = -1$. A dipole source is placed at $\mathbf{r}^{\prime}= (0.1,
-0.2, -0.5)$ and oriented along $\boldsymbol{\alpha}^{\prime}= (1/2, 1/2,
1/\sqrt{2})$ and fields are computed for $-5 \leq x \leq5$ and $-3 \leq z
\leq3$ for a fixed $y = 1.0$ with $\varepsilon_{1} = 1$, $\varepsilon_{2} =
2$, $\varepsilon_{3} = 4$, and $\lambda= 1$.}%
\label{three_layer_field_middle}%
\end{figure}

\begin{figure}[t]
\centering  \includegraphics[width=5.1in]{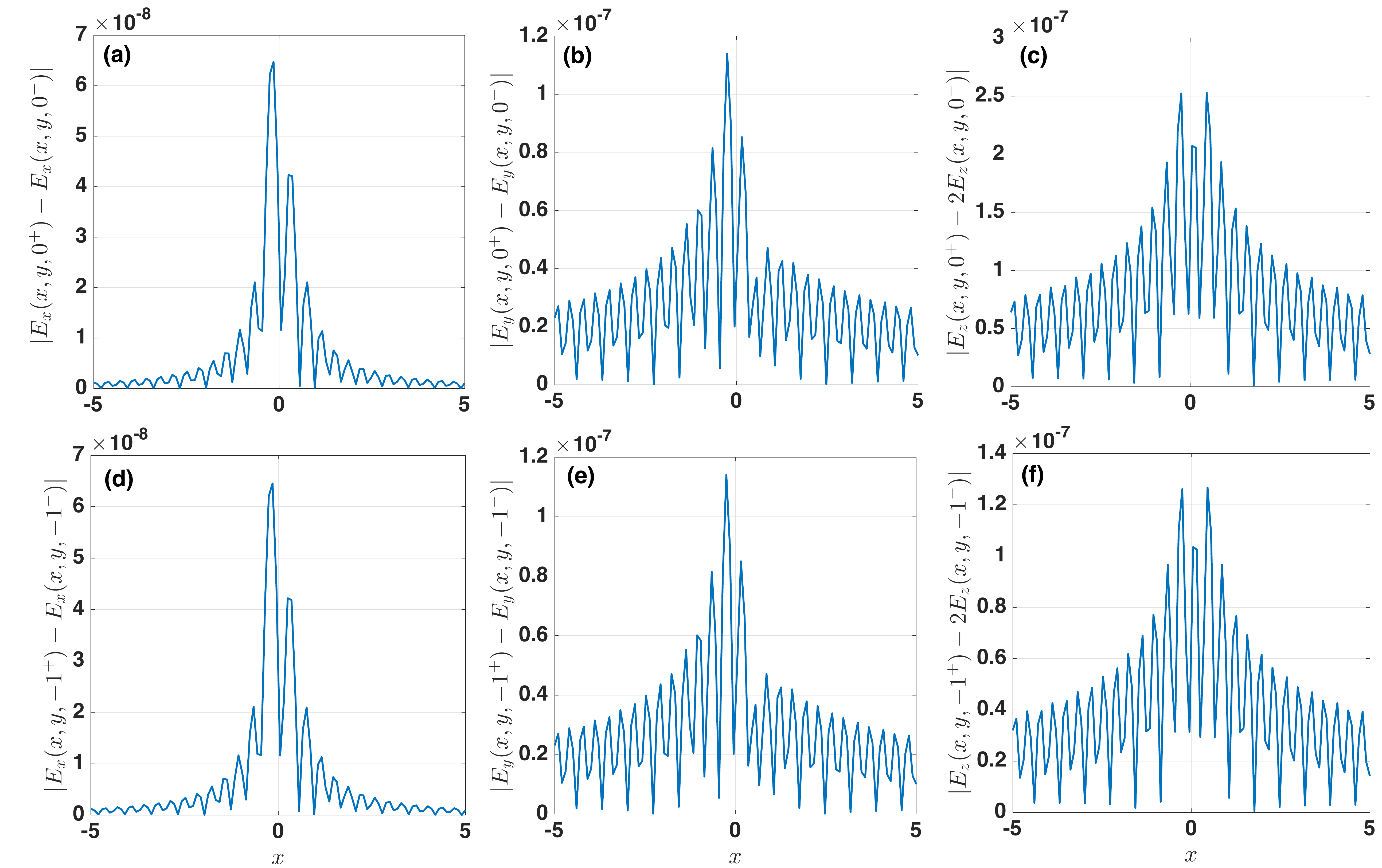}
\caption{Continuity of electric fields at both interfaces at $z = 0$ and $z =
-1$. (a) $|E_{x}(x,y,0^{+}) -E_{x}(x,y,0^{-})|$, (b) $|E_{y}(x,y,0^{+})
-E_{y}(x,y,0^{-})|$, (c) $|E_{z}(x,y,0^{+}) - 2E_{z}(x,y,0^{-})|$, (d)
$|E_{x}(x,y,-1^{+}) -E_{x}(x,y,-1^{-})|$, (e) $|E_{y}(x,y,-1^{+})
-E_{y}(x,y,-1^{-})|$, (f) $|E_{z}(x,y,-1^{+}) - 2E_{z}(x,y,-1^{-})|$. A dipole
source is placed at $\mathbf{r}^{\prime}= (0.1, -0.2, -0.5)$ and oriented
along $\boldsymbol{\alpha}^{\prime}= (1/2, 1/2, 1/\sqrt{2})$ and fields are
computed for $-5 \leq x \leq5$ for a fixed $y = 1.0$ at the layer interface $z
= 0$ and $z = -1$ and with $\varepsilon_{1} = 1$, $\varepsilon_{2} = 4$,
$\varepsilon_{3} = 1.1$, and $\lambda= 1$.}%
\label{three_layer_middle_error}%
\end{figure}

\subsubsection{Numerical results}

The Green's function is computed when the source is placed in the second
layer. Consider a three-layer structure defined by two interfaces located at
$z = 0$ and $z = -1$. The relative permittivity in each layer is $\varepsilon_{1}$ = 1,
$\varepsilon_{2}$ = 2, $\varepsilon_{3}$ = 4 and a dipole source is placed at
$\mathbf{r}^{\prime}= (0.1, -0.2, -0.5)$ oriented along $\boldsymbol{\hat{\alpha}}^{\prime}= (1/2, 1/2, 1/\sqrt{2})$ in the second layer. The relative permeability
$\{ \mu_{i} \}_{i = 1}^{3}$ is assumed to be $1$ in all layers. The wavelength
is set as $\lambda= 1$. In Fig. \ref{three_layer_field_middle}, all the components of total electric
field are plotted over $-5 \leq x \leq5$ and $-3 \leq z \leq3$ for a fixed $y
= 1.0$. The continuity of the fields are checked at both interfaces $z = 0$
and $z = -1$ in Fig. \ref{three_layer_middle_error}. In all components, about
$10^{-7}$ is achieved.

\section{Conclusion}

The electric field dyadic Green's function for a two- and three-layer
structure in 3-D are presented. The two-layer Green's function is simpler than
the one in Ref. \cite{chew1999} and uses one less Sommerfeld integral. An
adaptive generalized quadrature rule is applied to Sommerfeld integral to
obtain very high accuracy. Therefore, the proposed method is more accurate and
fast. Also it can be easily extended multi-layered media without any
modification except replacing the reflection and transmission coefficient. As
an example, a three-layer Green's function is presented to show the easy
extension to multi-layered media. The singular part is naturally separated as
a primary field that is the free-space Green's function. Therefore, the
Green's function is readily applicable to integral equation methods. The
Lippmann-Schwinger type volume integral equation used for the free space in
Ref. \cite{vie_jcp} is being modified with the new Green's function to study
many scatterers embedded in layered media.

As a relevant research issue, a fast solver will be developed using the
derived formulas for the Green's function for large-scale problems. Either a
new fast multipole method type method \cite{rokh90, lapFMM, scichina} or a
preconditioner \cite{precond_ying} based method could be considered to
accelerate an iterative matrix solver.

\section*{Acknowledgement}

This work was supported by a grant from the Simons Foundation (\#404499, Min
Hyung Cho) and W. Cai is supported by US Army Research Office (Grant No.
W911NF-14-1-0297) and US NSF (Grant No. DMS-1619713). The authors also like to
thank Dr. William Beck from Army Research Laboratory for helpful discussions during this work.

\appendix

\section{Bessel identities}

\label{besselidentity} A derivation of the dyadic Green's function in
multi-layered media is very tedious but it is required for a proper
implementation. The Bessel identities play a key role in the derivation and they are based on a integral representation of the
Bessel function and recurrence relation (See Ref. \cite{handbook}), namely,
\begin{equation}
J_{n}(z) e^{in\theta} = \frac{1}{2\pi} \int_{0}^{2\pi} e^{iz \cos{(\phi
-\theta)}+in\phi- in\frac{\pi}{2}} d\phi,
\end{equation}
\begin{equation}
J_{n+2}(z) = \frac{n+2}{z}J_{n+1}(z)-J_{n}(z),
\end{equation}
respectively. For convenience, the most often used identities are listed in
the following
\begin{align}
\int_{0}^{2\pi} e^{iz \cos{(\phi-\theta)}}~d\phi & = 2\pi J_{0}(z),\label{id1}%
\\
\int_{0}^{2\pi} e^{iz \cos{(\phi-\theta)}}\cos{\phi}~d\phi & = 2\pi i
J_{1}(z)\cos{\theta},\label{id2}\\
\int_{0}^{2\pi} e^{iz \cos{(\phi-\theta)}}\sin{\phi}~d\phi & = 2\pi i
J_{1}(z)\sin{\theta},\label{id3}\\
\int_{0}^{2\pi} e^{iz \cos{(\phi-\theta)}}\cos{2\phi}~d\phi & = -2\pi
J_{2}(z)\cos{2\theta},\label{id4}\\
\int_{0}^{2\pi} e^{iz \cos{(\phi-\theta)}}\sin{2\phi}~d\phi & = -2\pi
J_{2}(z)\sin{2\theta},\label{id5}\\
\int_{0}^{2\pi} e^{iz \cos{(\phi-\theta)}}\cos^{2}{\phi}~d\phi & = \pi
J_{0}(z)-\pi J_{2}(z)\cos{2\theta},\\
\int_{0}^{2\pi} e^{iz \cos{(\phi-\theta)}}\sin^{2}{\phi}~d\phi & = \pi
J_{0}(z)+\pi J_{2}(z)\cos{2\theta}.%
\end{align}


\bibliographystyle{elsarticle-num}
\bibliography{MinHyung}





\end{document}